\documentclass[twocolumn,superscriptaddress,amsmath,amssymb,floatfix,aps,prl,showpacs]{revtex4-1}
\usepackage{amsmath}    
\usepackage{graphicx}   
\usepackage{verbatim}   
\usepackage{color}      
\usepackage{subfigure}  
\usepackage{hyperref}   
\usepackage{mathrsfs}
\usepackage{amsfonts}
\usepackage{graphicx}   
\usepackage{verbatim}   
\usepackage{color}      
\usepackage{subfigure}  
\usepackage{hyphenat}
\usepackage[normalem]{ulem}
\usepackage{array}

\usepackage{color}      
\usepackage{soul}
\usepackage{epstopdf}
\newcommand{\teo}[1]{{\color{black} #1}}
\newcommand{\prtw}[1]{{\color{black} #1}}
\newcommand{\teob}[1]{{\color{black} #1}}

\begin{document}

\title{Fundamental costs in the production and destruction of persistent polymer copies} \author{Thomas E. Ouldridge}
\affiliation{Department of Bioengineering, Imperial College London,
  London, SW7 2AZ, UK} 
\email{t.ouldridge@imperial.ac.uk}
\author{Pieter Rein ten Wolde} \affiliation{FOM
  Institute AMOLF, Science Park 104, 1098 XE Amsterdam, The
  Netherlands}

\begin{abstract}
\teo{Producing a polymer copy of a polymer template is central to biology, and effective copies must persist after template separation.} We show that  this separation has three fundamental thermodynamic effects. Firstly, polymer-template interactions \teo{do not contribute to overall reaction thermodynamics} and hence cannot drive the process. \teo{Secondly, the equilibrium state of the copied polymer is template-independent and so additional work is required to provide specificity}. Finally, the mixing of copies from distinct templates makes correlations between template and copy sequences unexploitable, combining with copying inaccuracy to reduce the free energy stored in a polymer ensemble. \teo{These basic  principles set limits on the underlying costs and resource requirements, and suggest design principles, for autonomous copying and replication in biological and synthetic systems.}
\end{abstract}

\maketitle


Polymer copying is ubiquitous in living cells, occurring during replication, transcription and translation. These processes yield two physically separated, sequence-related polymers from a single input\cite{Alberts2002}. Previous work has addressed 
 the growth of a copy attached to a template \cite{Bennett1979, Cady2009,Andrieux2008,Sartori2013,Sartori2015}, but these  processes of templated self-assembly or templated polymerization do not directly produce {\it persistent} copies that are physically separated from their templates. Notably, whilst  templated self-assembly has been realized in  autonomous artificial  systems \cite{Luther1998,Kim2015,Sadownik2016,Schulman2012}, subsequent separation of copies without external manipulation has not. Similarly,
a tendency to remain template-bound has inhibited the generalization  to polymers   \cite{Orgel1992} of autocatylatic dimerization \cite{Sievers1994, Vidonne2009, Lincoln2009}. These difficulties emphasize that producing persistent copies involves more than just templated self-assembly.

\teo{We consider the fundamental thermodynamics of producing
  persistent copies, identifying the minimal work input through
  non-equilibrium free-energy changes. Eventual separation implies
  that, unlike in templated self-assembly, copy-template interactions
  cannot reduce the work required to produce a persistent
  copy. Moreover, a more accurate copy, which is more similar to
  its template, has a higher free energy and requires more work to
  create it.  \prtw{Different persistent copies produced from distinct
    templates can mix, however,} rendering copy-template sequence
  correlations unexploitable and 
  reducing {the minimal work required for copying}. \prtw{Our analysis provides
    fundamental bounds on the efficiency of cellular recylcing
    networks and on the resource requirements for natural
    and artifical copying systems, while suggesting design principles for
    (autonomous) copying systems.}}

\begin{figure}
\centering
\includegraphics[width=85mm]{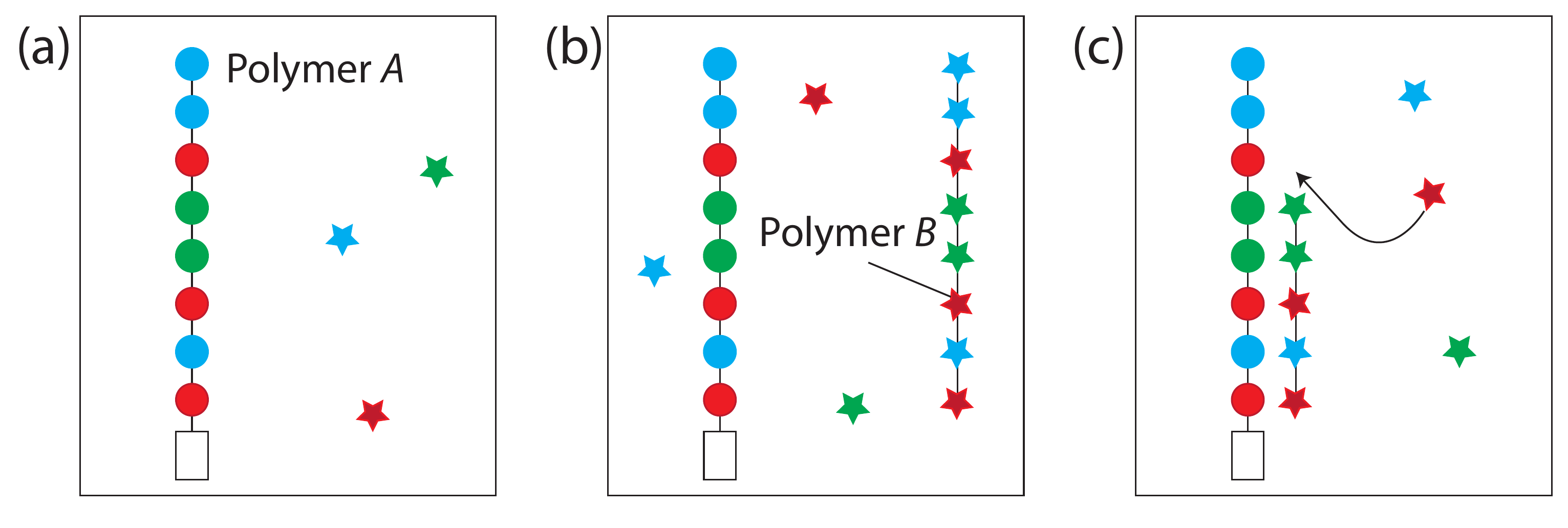}
\caption{Persistent copying of a polymer. (a) The initial state, with a polymer of class $A$ monomers. The final state, with $A$ unchanged and a second polymer of class $B$ monomers. The copying protocol induces a sequence of $B$ that is a copy of $A$, but no direct interactions are present in the final state. (c) Possible intermediate state, in which direct binding allows the sequence of $A$ to influence the sequence of $B$ as it grows. \label{fig:init_final}}
\end{figure}

We consider a polymer template of $N$ monomers, with $m$
different monomer types of class $A$, which might be
deoxyribonucleotides with $m=4$. We label the whole polymer $A$, with
a sequence vector ${\bf a}$ (Fig.\,\ref{fig:init_final}\,(a)).
We then grow a polymer $B$ from monomers of class $B$ of $m$ different
types, with a sequence ${\bf b}$ that is a copy of ${\bf a}$. After
the protocol, $B$ is physically separated from $A$, as illustrated in
Fig. \,\ref{fig:init_final}\,(b).  The sequences ${\bf a}$ and ${\bf
  b}$, and whether or not the two polymers are bound, together define
a biochemical macrostate ${\bf y}$ of system $Y$.  \teo{For a fixed
  sequence ${\bf a}$, the set of possible macrostates is then
  $\mathcal{B} = (\emptyset,\{1^*\},\{2\},\{2^*\}...)$, where
  $\emptyset$ indicates no $B$ polymer is present and no $B$
  monomers are bound to $A$, $\{n^*\}$ includes macrostates of all
  possible sequences of $B$ of length $n$ when bound to $A$, and
  $\{n\}$ includes all sequences of $B$ of length $n$
  when unbound.}

For our simple protocols we can work at the macrostate level.  The work required  to convert  $Y$ from  a macrostate distribution $\phi({\bf y})$ to  $\phi^\prime({\bf y})$ is bounded by the non-equilibrium free energy difference \cite{Esposito2011,Parrondo2015}:
$\langle w_{\phi \rightarrow \phi^\prime} \rangle \geq \mathscr{F}[\phi^\prime({\bf y})] - \mathscr{F}[\phi({\bf y})]$, 
with the equality holding for a reversible process, 
and 
$
\mathscr{F}[\phi({\bf y})] =  \mathcal{U}[\phi({\bf y})]- T \mathcal{S}[\phi({\bf y}))]
$.
Here,
$
\mathcal{U}[\phi({\bf y})]= \sum_{{\bf y}} \phi({\bf y}) U({\bf y}), 
$
and 
$
\mathcal{S}[\phi({\bf y})]= \sum_{{\bf y}} \phi({\bf y}) S({\bf y})  -k_B \sum_{{\bf y}} \phi({\bf y}) \ln \phi({\bf y}),
$
are the average energy and entropy, respectively. \teo{ The average chemical free energy $\mathcal{F}[\phi({\bf y})] =\sum_{\bf y} \phi({\bf y}) F({\bf y})= \sum_{\bf y} \phi({\bf y})( U({\bf y}) - TS({\bf y}))$ incorporates the chemical energy and entropy of  implicit microscopic degrees of freedom; the additional term $\mathcal{H}(Y) = \mathcal{H}[\phi({\bf y})] = -k_B \sum_{{\bf y}} \phi({\bf y}) \ln \phi({\bf y})$ is the Shannon entropy of the macrostate distribution.}

\teo{For our protocols, $A$ is initially drawn from a sequence distribution $\phi({\bf a})=p({\bf a})$, and $B$ is absent (state $\emptyset$). At the end of the protocol, $A$ is unchanged, but a persistent copy $B$ is created with a sequence drawn from $\phi({\bf b}|{\bf a})= p_f({\bf b}|{\bf a})$.} The absence of $A$-$B$ interactions  in the initial and final  states
implies that the chemical free energy is a sum over separate contributions from $A$ and $B$: $\mathcal{F}[\phi({\bf a},{\bf b})] = \mathcal{F}_A[\phi({\bf a})] + \mathcal{F}_B[\phi({\bf b})]$, with $\phi({\bf b}) = \sum_{\bf a}\phi({\bf a}) \phi({\bf b}|{\bf a})$. However, the details of copying will generate sequence correlations (Fig. \,\ref{fig:init_final}\,(c)), so the sequence entropy is not additive: $\mathcal{H}(A,B) =  \mathcal{H}(A) +\mathcal{H}(B|A) = \mathcal{H}(A)+\mathcal{H}(B) - k_{\rm B}T \mathcal{I}(A;B)$ \cite{Parrondo2015}. Here the conditional entropy $\mathcal{H}(B|A) = - \sum_{{\bf a},{\bf b}} \phi({\bf a}) \phi({\bf b}|{\bf a}) \ln \phi({\bf b}|{\bf a}) $ is the average sequence entropy of $B$ given $A$, and the mutual information $\mathcal{I}(A;B) =  \sum_{{\bf a},{\bf b}}  \phi({\bf a}) \phi({\bf b}|{\bf a}) \ln \phi({\bf b}|{\bf a}) /\phi({\bf b})$ is the reduction in $\mathcal{H}(B)$ given knowledge of  $A$. Since $\phi({\bf a})=p({\bf a})$ is unchanged by the protocol, and $ \mathcal{H}[p_0({\bf b})]=0$ for the initial $B$-distribution $p_0({\bf b})$, the reversible work  is   
\begin{equation}
\langle w^f \rangle_{\rm rev} =  \mathcal{F}_B[p_f({\bf b})] - \mathcal{F}_B[p_0({\bf b})] - T \mathcal{H}_f(B)+k_{\rm B}T \mathcal I_f(A;B),
\label{eq:rev_work}
\end{equation}
 with $\mathcal{H}_f(B) = \mathcal{H}[p_f({\bf b})]$ and  $\mathcal{I}_f(A;B) = \mathcal{I}[p_f({\bf b}|{\bf a}), p({\bf a})]$. \teo{Setting $\mathcal{F}_B[p_0({\bf b})] =0$ would be a valid normalisation.}

 \teo{Previous studies on templated self-assembly have shown that
   favorable $A$-$B$ interactions reduce the work required to
   \prtw{assemble a polymer $B$ on a template $A$} \cite{Bennett1979,
     Andrieux2008, Sartori2013, Sartori2015}.  Moreover, the presence of these
   interactions influences the equilibrium state of \prtw{the $B$
     polymer,  not only reducing the minimal work to
     grow a specific (desired) sequence, but also} \teob{providing a thermodynamic bias towards that sequence}
\cite{Andrieux2008, Sartori2013, Sartori2015}.  By
   contrast, the absence of $A$-$B$ interactions after copy-separation implies that the final free energy in a persistent copy process depends solely
   on interactions \prtw{{\em within}} $B$, and not \prtw{with} $A$.
   Thus both the reversible work $\langle w^f \rangle_{\rm rev}$ and
   the equillibrium distribution $p_{\bar{N}}^{\rm eq}({\bf b})$ that
   minimizes $ \mathcal{F}_B[\phi({\bf b})] - T \mathcal{H}(B)$ for an
   average length $\bar{N}$ are $A$-independent. Transitory binding
   during copying can neither reduce the overall work of copying, nor
   the relative cost of accurate \prtw{versus inaccurate copying}.  Indeed, a protocol producing
   a template-specific $\phi({\bf b}|{\bf a})=p_f({\bf b}|{\bf a})$
   always requires more work than one yielding a template-independent
   equilibrium distribution with the same average length $\bar{N}_f$,
   $\phi({\bf b}|{\bf a}) = p_{\bar{N}_f}^{\rm eq}({\bf b})$:}
\begin{align}
\langle w^f \rangle_{\rm rev} -  \langle w^{\rm eq}_{\bar{N}_f}\rangle_{\rm rev} =   \mathcal{F}_B[p_f({\bf b})] - \mathcal{F}_B[p_{\bar{N}_f}^{\rm eq}({\bf b})]  \label{eq:delta w} \\
  + T(\mathcal{H}^{\rm eq}_{\bar{N}_f}(B)-\mathcal{H}_f(B) ) +k_{\rm B}T \mathcal I_f(A;B) \geq 0. \nonumber
\end{align}
\teo{Here, we have used  $\mathcal{I}_{\bar{N}_f}^{\rm eq}(A;B) =0$ for independent  $A$ and $B$. The inequality follows from} $\langle w^f\rangle_{\rm rev} =   \mathcal{F}_B[p_f({\bf b})] - T \mathcal{H}_f(B) + k_{\rm B}T \mathcal I_f(A;B) \geq  \mathcal{F}_B[p_f({\bf b})] - T \mathcal{H}_f(B) \geq   \mathcal{F}_B[p_{\bar{N}_f}^{\rm eq}({\bf b})] - T \mathcal{H}_{\bar{N}}^{\rm eq}(B) = \langle w^{\rm eq}_{\bar{N}_f}\rangle_{\rm rev}  $. \teo{The lowest-cost output is template-independent, with sequences drawn from $p_{\bar{N}}^{\rm eq}({\bf b})$.  Template-specific  persistent copies {\it necessarily} require more work because specific copies {\it necessarily} have higher free energy, unlike in templated self-assembly. }

\teo{Neither $\langle w^{\rm eq}_{\bar{N}_f} \rangle_{\rm rev}$, nor
  $\langle w^f \rangle_{\rm rev} - \langle w^{\rm eq}_{\bar{N}_f}
  \rangle_{\rm rev}$, are dissipated, but stored in the final free
  energy. Three terms contribute to $\langle w^f \rangle_{\rm rev} -
  \langle w^{\rm eq}_{\bar{N}_f} \rangle_{\rm rev}$: a difference in
  chemical bonds within $B$, $ \mathcal{F}_B[p_f({\bf b})] -
  \mathcal{F}_B[p_{\bar{N}_f}^{\rm eq}({\bf b})]$; a difference in
  sequence entropy $\mathcal{H}^{\rm eq}_{\bar{N}_f}(B)
  -\mathcal{H}_f(B) $; and $k_{\rm B}T \mathcal I_f(A;B)$,  reflecting the free energy stored in correlations
  \cite{Horowitz2013,Parrondo2015,Ouldridge:2015vi,McGrath2016}, since non-interacting
  $A$ and $B$ are statistically independent in equilibrium.} 
 \teo{The first two
  terms can be individualy positive or negative, but the third, and
  the sum, are necessarily non-negative. Combining the final two terms
  gives a single copying accuracy contribution, $T(\mathcal{H}^{\rm
    eq}_{\bar{N}}(B)-\mathcal{H}_f(B) ) +k_{\rm B}T \mathcal I_f(A;B)
  = T(\mathcal{H}^{\rm eq}_{\bar{N}}(B)-\mathcal{H}_f(B|A)
  )$. Perfect
  copying, with $\mathcal{H}_f(B|A)=0$, has a large cost.}

\teo{Despite not being dissipated, the minimal work required for accurate
  copying has implications for optimal replication. England used the
  total entropy increase as the replication cost, bounding it by
  the logarithm of the ratio of the replicator's birth and death rates
  \cite{England:2013ed}. Since this ratio can approach unity at an
  arbitrary net replication rate, there is no apparent minimal cost
  per replication.  \prtw{However}, replication accuracy is absent in
    this analysis.  \prtw{Yet,} replicators must make persistent
      copies, and \prtw{our analysis shows} that copy accuracy bounds
      the chemical work or resources required. Even if replication is
      reversible, generating zero total entropy, these resources
      cannot be recovered by the parent without reversing the copy and hence
      destroying the offspring. Thus increased accuracy
      necessarily requires more  resources that could be used
      elsewhere, such as to produce more offspring. }

We illustrate a reversible copying  protocol  in 
Fig.\,
{fig:protocol}. We nucleate $B$ from a seed, to which an external force can be applied, and we manipulate the chemical potential of $B$-type monomers via a series of buffers \cite{Ouldridge:2015vi}. To  produce a single copy,  $B$ must only grow or shrink from its tip when in contact with $A$, and cannot grow beyond length $N =|{\bf a}|$; a catalyst could facilitate the desired reactions whilst keeping all others slow. We also assume that the $i^{\rm th}$ monomer in $A$ can only interact with the  $i^{\rm th}$ monomer in $B$. Though idealized, the system is thermodynamically valid since all reactions have  a microscopic reverse. 

\begin{figure}
\centering
\includegraphics[width=85mm]{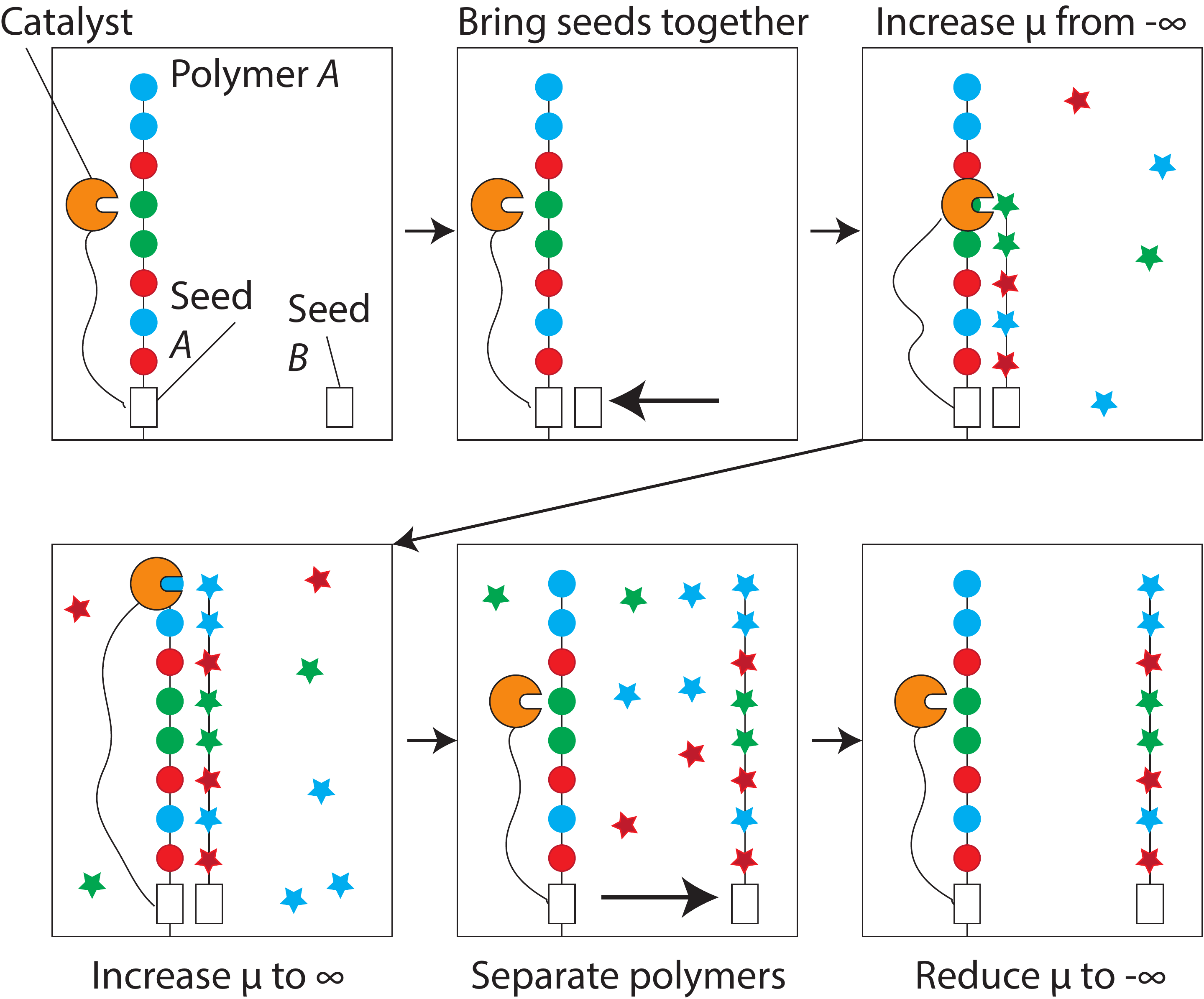}
\caption{A reversible protocol for persistent copying. Initially, seed $B$ is separate from $A$, with  $B$-monomers present at a low chemical potential, $\mu \rightarrow -\infty$. The external force brings seed $B$ into contact with $A$ quasi-statically, extracting work. The chemical potential $\mu$ of monomers is slowly raised, causing $B$ to grow. Eventually, $\mu \rightarrow \infty$ and  $|{\bf b}|=|{\bf a}|$. At this point, the external force  separates the two polymers quasistatically, doing work against the binding free energy. Finally, the chemical potential of monomers is returned to its initial value.  \label{fig:protocol}}
\end{figure}

Given $F({\bf b})$,  $\langle w^f \rangle_{\rm rev}$ is calculable. Let  the binding free energy of  seeds be $\Delta F_{\rm s}$, and assume that adding a monomer $x$ to an isolated $B$ changes the chemical free energy of $B$ by  $\Delta F_x$. When in contact with $A$, $\Delta F_x$ is modified by $\Delta F_{\rm c}$ for correct matches, and $\Delta F_{\rm nc}$ otherwise. 
Mechanical work   $\langle w_{\rm seed} \rangle = \Delta F_{\rm s} + C$  is extracted on bringing the seeds  into contact ($C$ reflects initial dilution). Chemical work is done during polymer growth, as the chemical potential of monomers is raised: 


\begin{align}
&\langle w_{\rm pol}({\bf a}) \rangle = -k_{\rm B}T \ln \\ 
&\left(\sum_{{\bf b},|{\bf b}|=n}  \prod_{x=1}^N {\rm e}^{\frac{-\Delta F_{b_x}}{k_{\rm B}T}} \left((1-\delta_{a_x b_x}){\rm e}^{\frac{-\Delta F_{\rm nc}}{k_{\rm B}T}} + \delta_{a_x b_x}{\rm e}^{\frac{-\Delta F_{\rm c}}{k_{\rm B}T}}\right) \right), \nonumber
\end{align}
as shown in  Section 1 of Ref.\,\cite{SI}.
Separation  requires mechanical work
$
\langle w_{\rm sep}({\bf a})\rangle = -\Delta F_{\rm s} - C - \Delta F_{AB}({\bf a}).
$
Here, $\Delta F_{AB}({\bf a})$ is the average contribution to the chemical free energy of polymerization from the $A$-$B$ interaction,
\begin{equation}
{\Delta F_{AB}({\bf a})}=   \sum_{{\bf b}} p_f({\bf b}|{\bf a}) \sum_{x=1}^{|{\bf b}|} \left((1- \delta_{a_x b_x}) F_{\rm nc}  + \delta_{a_x b_x} F_{\rm c} \right). 
\end{equation}
The double-edged role of attractive interactions between $A$ and $B$ (negative $F_{\rm c}$ and $F_{\rm nc}$) is evident. They reduce ${\langle w_{\rm pol}({\bf a})\rangle }$, but  provide a corresponding increase in $\langle w_{\rm sep}({\bf a})\rangle$.
Summing $\langle w_{\rm pol} ({\bf a})\rangle$, $\langle w_{\rm sep}({\bf a}) \rangle$ and $\langle w_{\rm seed} \rangle$, and averaging over $p({\bf a})$ (Section 2 of Ref.\,\cite{SI}), yields 
\begin{equation}
\langle w^f \rangle = k_{\rm B} T  \sum_{{\bf b}} p_f({\bf b}) \sum_{n=1}^{|{\bf b}|}  \Delta F_{b_n} -T \mathcal{H}_f(B) + k_{\rm B}T \mathcal{I}_f(A;B),
 \label{eq:work_final}
\end{equation}
in which the dependence on $F_{\rm c}$ and $F_{\rm nc}$ has canceled. The first term is  $\mathcal{F}_B[p_f({\bf b}) ] -\mathcal{F}_B[p_0({\bf b})]$, the change in average chemical free energy. Thus
Eq.\,\ref{eq:work_final} matches Eq.\,\ref{eq:rev_work},
confirming that the protocol is  reversible. Indeed, reversing the protocol recovers $\langle w^f \rangle$ and restores the initial state.
\teo{A finite  growth rate or non-equilibrium proofreading during the polymerization stage, as considered in Refs.~\cite{
Bennett1979, Cady2009,Andrieux2008,Sartori2013,Sartori2015}, would lead to an increase in work over the minimum required by the output distribution $p_f({\bf b}|{\bf a})$, $\langle w^f \rangle  > \langle w^f \rangle_{\rm rev}$.}

\teo{Cells produce different persistent RNA and protein molecules from
  multiple distinct templates, and these copies subsequently
    mix. Motivated by this observation} we now consider an ideal mixture of  $M$
persistent copies of a given set of templates. The copy macrostate is
now specified by the numbers of each sequence present $\{M_{\bf b}\}$,
with a distribution $\phi(\{M_{\bf b}\})$. The copies have free
energy
\begin{align}
\mathscr{F}[\phi(\{M_{\bf b}\})] = -k_{\rm B} T \sum_{\{M_{\bf b}\}} \phi(\{M_{\bf b}\}) \ln \prod_{\bf b} \frac{Z_{\bf b}^{M_{\bf b}}}{M_{\bf b}!}  \label{eq:multiF1}\\
+ k_{\rm B} T  \sum_{\{M_{\bf b}\}} \phi(\{M_{\bf b}\}) \ln \phi(\{M_{\bf b}\}). \nonumber
\end{align}
The first term is the average chemical free energy  $\mathcal{F}_B[ \phi(\{M_{\bf b}\})] = \sum_{\{M_{\bf b}\}} \phi(\{M_{\bf b}\}) F(\{M_{\bf b}\})$, and the second the macrostate entropy $-k_{\rm B}T\mathcal{H}[ \phi(\{M_{\bf b}\})]$. Here $F(\{M_{\bf b}\}) = -k_{\rm B} T  \ln \prod_{\bf b} {Z_{\bf b}^{M_{\bf b}}}/{M_{\bf b}!} $ is the standard expression for dilute solutes  with $ -k_{\rm B} T \ln Z_{\bf b}$ the chemical free energy of an isolated polymer \cite{Huang1987}. For the simple model considered previously, $Z_{\bf b} = Z_0 \prod_{x=1}^{|{\bf b}|} {\rm e}^{{-\Delta F_{b_x}}/{k_{\rm B}T} }$, with $-k_{\rm B}T \ln Z_0$ the free energy of an isolated seed.

To compare with our previous result, let each copied template be drawn from $p({\bf a})$ (for an alternative, see Section 3 of Ref.\,\cite{SI}), giving $p_f({\bf b}) = \sum_{\bf a} p({\bf a}) p_f({\bf b}| {\bf a})$. In this case,  $\phi(\{M_{\bf b}\}) = M!\prod_{\bf b}  p_f({\bf b})^{M_{\bf b}}/M_{\bf b}! $. Substituting  into Eq.\,\ref{eq:multiF1}  and using $\sum_{\{M_{\bf b}\}} \phi(\{M_{\bf b}\}) M_{\bf b} = \langle M_{\bf b} \rangle = M p_f({\bf b})$, we obtain
\begin{align}
\mathscr{F}[\phi(\{M_{\bf b}\})] = -k_{\rm B} T M \sum_{\bf b} p_f({\bf b}) \ln  Z_{\bf b} \label{eq:multiF3} \\
+ k_{\rm B} T M \sum_{\bf b} p_f({\bf b} ) \ln  p_f({\bf b} ) +  k_{\rm B} T \ln  M!. \nonumber
\end{align}
The first term is the average chemical free energy of $M$ isolated copies, $M\mathcal{F}_B[p_f({\bf b})]$, and the second the entropy $ -T M\mathcal H_f(B)$. The third  term is independent of the copying details. As before,   $\mathscr{F}$ (and hence required work) is template-independent, and  is minimal for $p_f({\bf b}) = p_{\bar{N}_f}^{\rm eq}({\bf b})$. Thus for many copies, $(\langle W^f\rangle_{\rm rev} - \langle W^{\rm eq}_{\bar{N_f}}\rangle_{\rm rev})/M = \mathcal{F}_B[p_f({\bf b})] - \mathcal{F}_B[p_{\bar{N}_f}^{\rm eq}({\bf b})] + T\mathcal H^{\rm eq}_{\bar{N}_f}(B) -T  \mathcal H_f(B)$. Absent is the $k_{\rm B}T \mathcal{I}_f(A;B)\geq 0$ copy-template correlation term that is present in the single copy case (Eq.\,\ref{eq:delta w}).
Only the template-averaged distribution $p_f({\bf b})$ matters, and differences between copies of distinct templates are irrelevant.

Correlations do not contribute to $\mathscr{F}$ in 
the multi-copy case
due to mixing.
When pairs of correlated non-interacting molecules are
identifiable, as when copy-template pairs are isolated, the
correlations are exploitable \cite{McGrath2016}. Once mixed, however,
templates cannot be matched to copies {\it a priori}, and correlations
cannot be leveraged. The stored free energy is no higher than if each
template gave a non-specific distribution $p_f({\bf b}|{\bf a}) =
p_f({\bf b})$. \teo{If all templates have the same sequence, mixing
  copies has no effect, and the free energy is unchanged. Indeed,
  $\mathcal{I}_f(A;B)=0$ in this case, since $\mathcal{H}(A)=0$, and
  hence $(\langle W^f\rangle_{\rm rev} - \langle W^{\rm
    eq}_{\bar{N_f}}\rangle_{\rm rev})/M=\langle w^f\rangle_{\rm rev} -
  \langle w^{\rm eq}_{\bar{N_f}}\rangle_{\rm rev}$}.

\begin{figure}
\centering
\includegraphics[width=42mm,angle=-90]{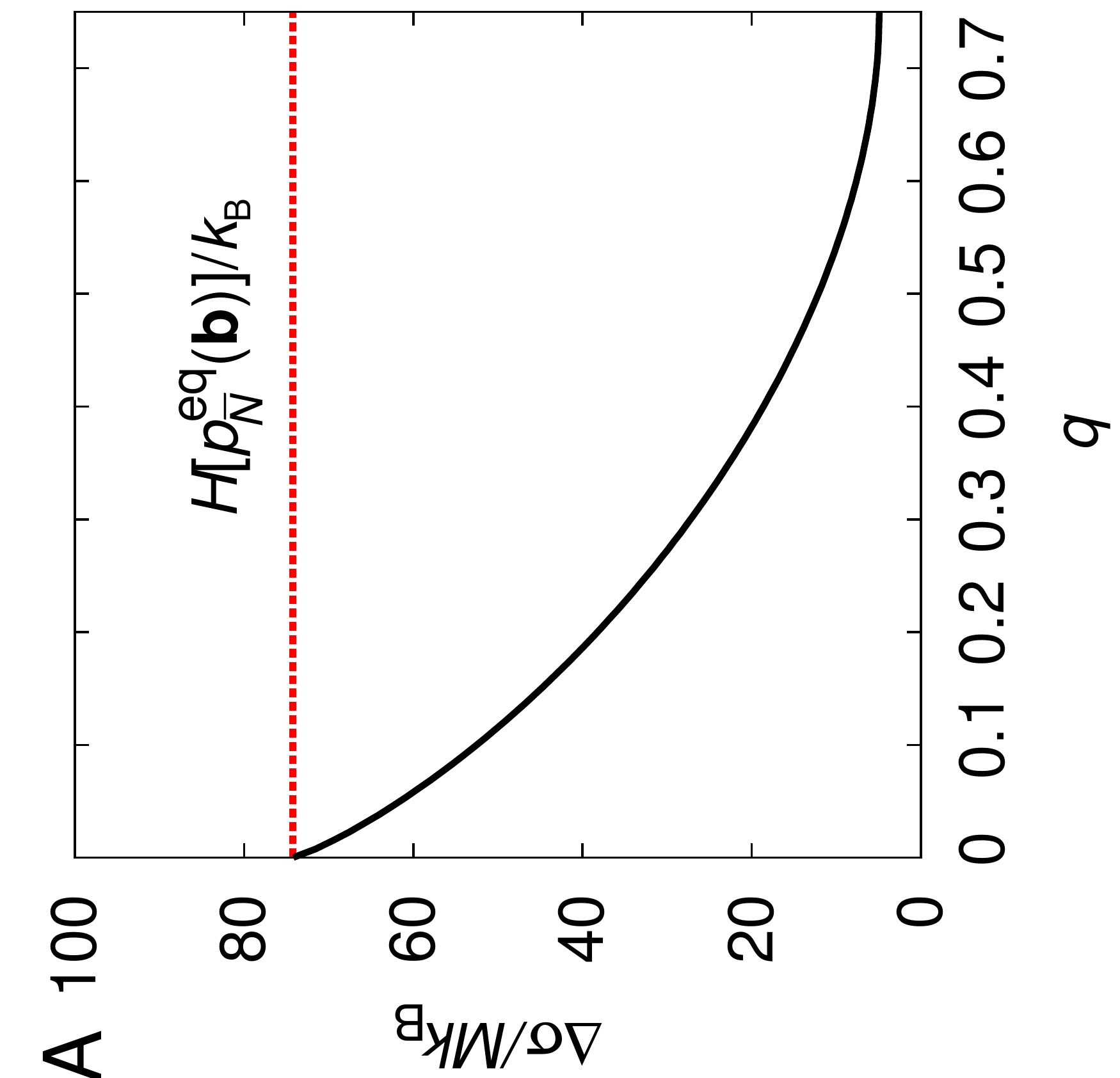}
\includegraphics[width=45mm,angle=-90]{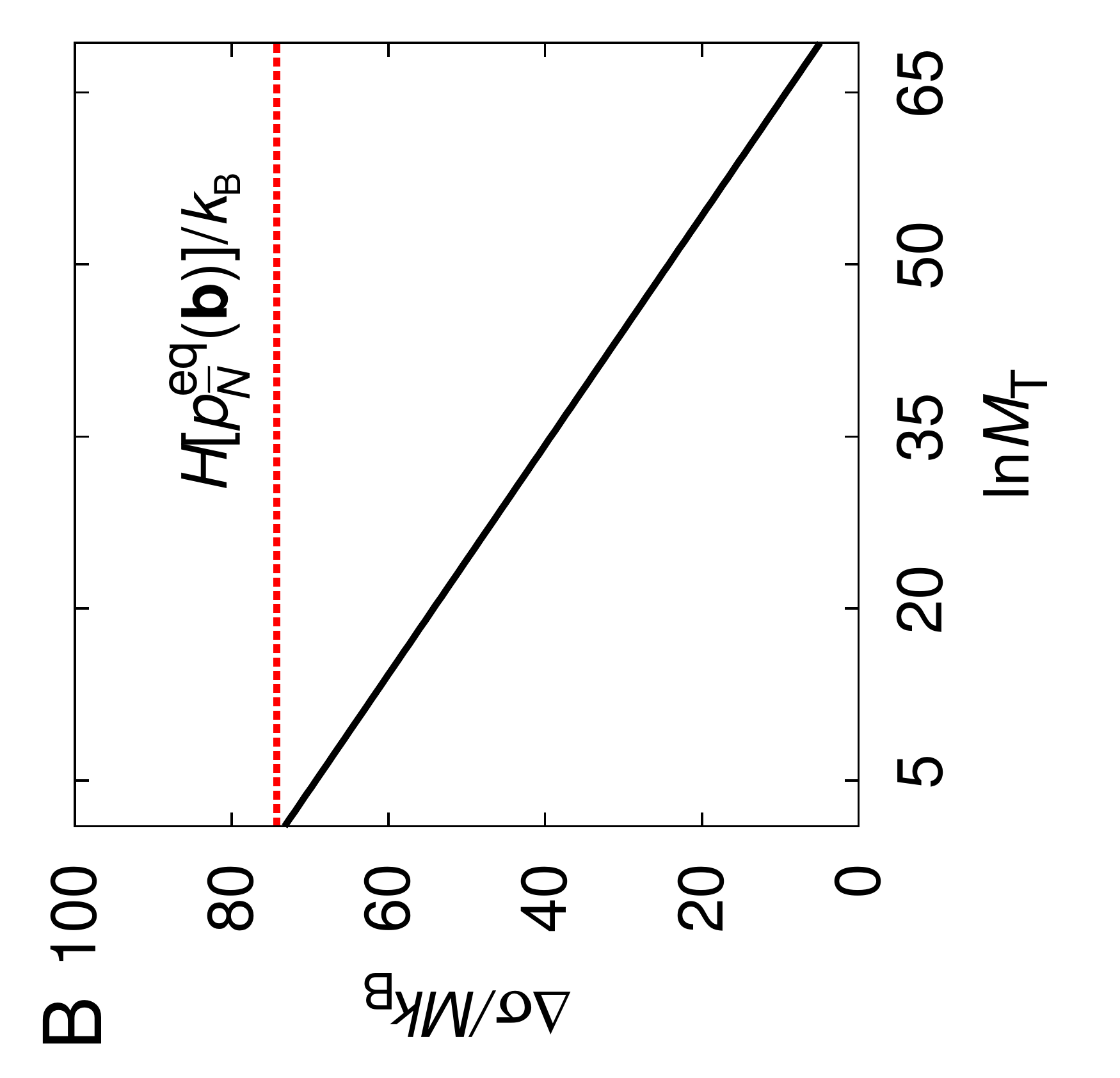}
\caption{Inaccurate copying and the presence of copies from multiple templates reduce the minimal entropy generation during non-specific depolymerization. We plot  $ \Delta \sigma =  \mathcal{H}[p^{\rm eq}_{\bar{N}_z}({\bf b})]  - \mathcal{H}[p_z(b)] $, the entropy generated by the non-specific depolymerization protocol discussed in the text when all monomers are equally stable within $B$. We consider an ensemble of polymers all within initial length $\bar{N} =50$ and with four distinct monomers ($m=4$). (a) All copies produced from a single template, with an error rate of $q \leq (m-1)/m$ per monomer. (b) Copies produced from with 100\% accuracy and equal probability from $M_{\rm T}$ distinct templates, with $1\leq M_{\rm T} \leq 4^{\bar{N}}$. Neither graph reaches zero because the initial ensembles always contain a single polymer length.
  \label{fig:depol_costs}}
\end{figure}

To reach the lower bound  $\langle W^f\rangle_{\rm rev}$ on the work to produce a mixed ensemble $\phi(\{M_{\bf b}\}) = M!\prod_{\bf b}  p_f({\bf b})^{M_{\bf b}}/M_{\bf b}! $, a process must  exploit the free energy released upon mixing -- we outline such a protocol in Section 4 of Ref.\,\cite{SI}. If, instead, mixing simply  occurred irreversibly after reversible copying,  the entropy of the universe would increase by the excess work $(\langle W^f \rangle - \Delta \mathscr{F}[\phi(\{M_{\bf b}\}) ])/T = k_B \mathcal{I}(A;B)$. 

\teo{Cells recycle RNA and proteins via irreversible non-specific
depolymerization pathways} \cite{Bennett1982}, rather than by measuring
sequences and depolymerizing with an appropriate template. \teo{In
  such {\em cyclic} operations, unlike replication, \prtw{total} entropy
  generation measures recycling inefficiency  and is the natural metric for cost. The entropy generated in
  depolymerization sets a lower bound on the cost of the entire
  cycle.}  \teo{Bennett claimed that template-free depolymerization
  would generate at least $kT \ln m$ of entropy per monomer
  depolymerised, with $m$ the number of distinct monomer types; other
  authors have found similar results \cite{Bennett1982, Andrieux2013,
    Gaspard2014, Gaspard2015,Gaspard2016}. However, these analyses
  consider a single initial  sequence
 and hence
  underestimate the initial polymer entropy by assuming it is
  zero \cite{Andrieux2013, Gaspard2014, Gaspard2015,Gaspard2016}. In
  reality the sequence entropy depends on the distribution of initial
  sequences, with a broader distribution implying a  greater initial
  entropy.}

For concreteness, consider the earlier model with $M$ polymers and a distribution of macrostates $\phi(\{M_{\bf b}\})=  M!\prod_{\bf b}  p_z({\bf b})^{M_{\bf b}}/M_{\bf b}!$.  To depolymerize non-specifically, we set $\mu= \bar{{\mu}}_z$ such that the average equilibrium length equals the average initial length $\bar{N}_z$ of polymers, and introduce catalysts that allow growth/shrinking. With this choice of $\mu$ there is no change in $\bar{N}$ when the catalysts are first introduced, and hence  no chemical work since the net number of monomers transferred from the buffer is zero. Nonetheless,  the distribution relaxes irreversibly to the equilibrium $\phi(\{M_{\bf b}\})=  M!\prod_{\bf b}  p^{\rm eq}_{\bar{N}_z}({\bf b})^{M_{\bf b}}/M_{\bf b}!$, generating entropy 
\begin{align}
T \Delta \sigma_{\rm relax} =- k_{\rm B} T M \sum_{\bf b}\left( p_z({\bf b}) - p^{\rm eq}_{\bar{N}_z}({\bf b})\right)\ln  {Z_{\bf b}}  \label{eq:entropy}\\
+ k_{\rm B} T M \mathcal{H}[p^{\rm eq}_{\bar{N}_z}({\bf b})]  - k_{\rm B} T M \mathcal{H}[p_z(\bf b)], \nonumber
\end{align}
using Eq.\,\ref{eq:multiF3}.
Any other choice of initial $\mu$ would generate more entropy through unbalanced growth or shrinking. On taking $\mu \rightarrow -\infty$, the polymers shrink reversibly to zero, meaning that $T \Delta \sigma_{\rm relax}=T\Delta \sigma$ is the total increase in the entropy of the universe during depolymerization. 

We verify this dissipation for a specific model in Section 5 of
Ref.\,\cite{SI}. For the special case \teo{in which $\Delta F_x$ is $x$-independent, $\ln Z_{\bf b}\propto |{\bf b}|$ and thus as}
 $\bar{N} \rightarrow \infty$, $T \Delta \sigma = k_{\rm B}T M
\bar{N}_z \ln m - k_{\rm B} T M \mathcal{H}[p_z({\bf b})]$,
generalizing Bennett's result \cite{Bennett1982} to a distribution of
input polymers. Thus the minimal entropy generation of non-specific
recycling  depends on the \teo{the details of the \prtw{preceding} production of persistent copies}
(Fig.\,\ref{fig:depol_costs}).
  Non-specific
depolymerisation is cheap if the polymers are drawn from a broad
distribution due to inaccurate copying and/or a broad distribution of
templates. For the biological case of high accuracy and a limited
number of templates, the effect of non-zero $\mathcal{H}[p_z({\bf
  b})]$ is small compared to $\bar{N}_z \ln m$.

\teo{ Our analysis uses free-energy calculations, and the resulting
  bounds can only be reached by quasistatic operations. Our optimal
  protocol is non-autonomous, involving external
  manipulation. Nonetheless, it provides insight into autonomous
  copying in natural and synthetic systems.  Firstly, our results
  allow a meaningful definition of the efficiency of polymer copying,
  by comparing the work done to $\langle w^f\rangle_{\rm rev}$. Our
  analysis and its bounds provide a framework for the thermodynamics
  of producing persistent polymer copies, like the Carnot cycle does
  for heat engines.  Recently, we have shown the relevance of a
  similar bound for the autonomous, finite-speed copying of a receptor
  by a biochemical network \cite{Ouldridge:2015vi}. 

  Secondly,  {our results reveal} fundamental differences between
  the optimal designs of copying networks and superficially similar
  self-assembling systems.  Autonomous templated self-assembly can
  occur accurately and reversibly due to the equilibrium thermodynamic
  bias provided by favorable interactions between the matching
  monomers \cite{Andrieux2008, Sartori2013, Sartori2015}. Indeed,
  quasi-reversible conditions are generally seen as optimal for
  self-assembly \cite{Wilber2007,Reinhardt2014}. \prtw{We show,
 however, that the minimal work to make persistent copies
    does not depend on template-copy interactions (Eq. \ref{eq:delta
      w}), which means that no equilibrium bias towards correct copying is possible. 
The fact that template-copy
  interactions are absent in the final state implies that these
  interactions can only provide specificity if they selectively
  stabilize the intermediate states of the copy process. For an
  autonomous and continuously-operating system, this means that the
  template must act as a catalyst,  providing specificity via
  kinetic discrimination (we discuss non-autonomous systems in Section S6 of Ref.\,\cite{SI}).
 Kinetic discrimination, however, requires that the system is
  driven out of thermodynamic equilibrium;}
 we therefore predict that autonomous networks
  producing persistent copies  {\em must}  be non-specific in the
  reversible limit, \prtw{as seen for templated self-assembly when
    discrimination is based on kinetics rather than thermodynamics}
  \cite{Bennett1979,Sartori2013}. Dissipation in natural copying systems is therefore
  not only necessary to provide enhanced accuracy through proofreading
  \cite{Hopfield:1974ij, Bennett1979, Sartori2015}, but to provide {\it any
  accuracy at all}. Synthetic copying networks should therefore be
  designed fundamentally differently from near-equilibrium
  self-assembling systems.}

\teo{ Finally, by highlighting the double-edged role of template-copy  interactions, which enhance accurate polymerization but  inhibit dissociation, our work draws attention to the differences  between the distinct mechanisms that cells employ for persistent copying.  Nature has two approaches. Viewing DNA  replication at the level of the single strands, a copy is grown in  contact with its template, and the cost of its separation is paid for after the copy is made in full (to enable the next  replication). By contrast, in transcription and translation, the  copy is only attached to the template by a handful of monomers at  any one time; as new monomers join, older ones detach from the  template. The importance of template-copy  separation in terms of function and underlying thermodynamics suggests that the unique characteristics of these two distinct mechanisms warrant further  consideration.  }

\emph{Acknowledgements}: 
TO was supported by a Royal Society University Research Fellowship.
 This work is part of the research programme of the
  Foundation for Fundamental Research on Matter (FOM), which is part
  of the Netherlands Organisation for Scientific Research
  (NWO).

\begin{widetext}
\section{S1. Derivation of $\langle w_{\rm pol} ({\bf a})\rangle$} 
The chemical potential of species $x$ is $\mu_x = \partial F_{\rm buffer} / \partial N_x$; for simplicity, we choose uniform $\mu_x=\mu$. Thus the free-energy change of the buffer due to monomer transfer from buffer to the polymer, leading to the growth of the polymer by one unit, is $\Delta F_{\rm buffer} = -\mu$, equivalent to the expenditure of $-\Delta F_{\rm buffer} = \mu$ of chemical work. During  polymerization, the buffers therefore perform an average work for a given template sequence ${\bf a}$ of
\begin{equation}
\langle w_{\rm pol} ({\bf a}) \rangle = \int_{-\infty}^{+\infty} {\rm d} \mu \, \mu \frac{{\rm d} \langle|{\bf b}| \rangle_{\bf a}}{{\rm d}\mu}, 
\end{equation}
where $\langle|{\bf b}| \rangle_{\bf a}$ is the expected length of $B$  given $\mu$ and ${\bf a}$. 
When attached to $A$ and at
 chemical potential $\mu$, the relative probability of a specific configuration ${\bf b}$ given ${\bf a}$ is
\begin{equation}
\frac{P({\bf b}|{\bf a})}{P(0|{\bf a})} = {\rm e}^{\frac{\mu |{\bf b}|}{k_{\rm B}T}} \prod_{x=1}^{|{\bf b}|}{\rm e}^{\frac{-\Delta F_{b_x}}{k_{\rm B}T}} ((1-\delta_{a_x b_x}){\rm e}^{\frac{-\Delta F_{\rm nc}}{k_{\rm B}T}} + \delta_{a_x b_x}{\rm e}^{\frac{-\Delta F_{\rm c}}{k_{\rm B}T}}).
\end{equation}
The relative probability of  $|{\bf b}|= n$ is thus
$
{P(|{\bf b}|=n|{\bf a})}/{P(|{\bf b}|=0|{\bf a})} = {\rm e}^{{ \mu n}/{k_{\rm B}T}} Q(n|{\bf a})
$, with 
\begin{equation}
Q(n|{\bf a})= \sum_{{\bf b},|{\bf b}|=n}  \prod_{x=1}^n {\rm e}^{\frac{-\Delta F_{b_x}}{k_{\rm B}T}} ((1-\delta_{a_x b_x}){\rm e}^{\frac{-\Delta F_{\rm nc}}{k_{\rm B}T}} + \delta_{a_x b_x}{\rm e}^{\frac{-\Delta F_{\rm c}}{k_{\rm B}T}}).
\end{equation}
We will simplify this expression before using it  in the integral for chemical work.
We introduce
$ \theta =  \beta \mu + (1/N)\ln Q(N|{\bf a})$, where $N=|{\bf a}|$. In terms of this variable,
\begin{equation}
\frac{P(|{\bf b}|=n|{\bf a})}{P(|{\bf b}|=0|{\bf a})} = {\rm e}^{\theta n} \frac{Q(n|{\bf a)}}{Q(N|{\bf a})^{1/N}}.
\end{equation}
 Thus the expectation of $|{\bf b}|$ given a specific ${\bf a}$ is
\begin{equation}
\langle |{\bf b}|(\theta) \rangle_{\bf a} = \frac{{\rm d }}{{\rm d \theta}} \ln {\sum_n {\rm e}^{\theta n} \frac{Q(n|{\bf a})}{ {Q(N|{\bf a})}^{n/N} }},
\end{equation}
Consequently, the work integral becomes
\begin{equation}
\frac{\langle w_{\rm pol}({\bf a})\rangle }{k_{\rm B}T}= \int_{-\infty}^{+\infty} {\rm d} \theta    \left(\theta - \frac{\ln Q(N|{\bf a})}{N} \right) \frac{{\rm d}^2}{{\rm d}\theta^2} \ln {\sum_n {\rm e}^{\theta n} \frac{Q(n|{\bf a})}{ {Q(N|{\bf a})}^{\frac{n}{N}} }}.
\label{eq:integral2}
\end{equation}
The term  ${\ln Q(N|{\bf a})}/{N}$ is constant within the integral. Using the fact that $\underset{\theta \rightarrow -\infty}{\lim} \langle |{\bf b}_{\bf a}|(\theta) \rangle =0$ and $\underset{\theta \rightarrow \infty}{\lim} \langle |{\bf b}_{\bf a}|(\theta)  \rangle =N$,
\begin{equation}
\frac{\langle w_{\rm pol}({\bf a})\rangle }{k_{\rm B}T}=  - {\ln Q(N|{\bf a})} + \int_{-\infty}^{+\infty} {\rm d} \theta    \theta\frac{{\rm d}^2}{{\rm d}\theta^2} \ln {\sum_n {\rm e}^{\theta n} \frac{Q(n|{\bf a})}{ {Q(N|{\bf a})}^{{n}/{N}} }}.
\end{equation}
The second term can be integrated by parts 
\begin{align}
\int_{-\infty}^{+\infty} {\rm d} \theta    \theta\frac{{\rm d}^2}{{\rm d}\theta^2} \ln {\sum_n {\rm e}^{\theta n} \frac{Q(n|{\bf a})}{ {Q(N|{\bf a})}^{{n}/{N}} }} = \left[  \theta  \frac{{\rm d}}{{\rm d}\theta} \ln {\sum_n {\rm e}^{\theta n} \frac{Q(n|{\bf a})}{ {Q(N|{\bf a})}^{n/N} }} \right]^{\infty}_{-\infty} 
 - \int_{-\infty}^{+\infty} {\rm d} \ln {\sum_n {\rm e}^{\theta n} \frac{Q(n|{\bf a})}{ {Q(N|{\bf a})}^{n/N} }}.
\label{eq:int3}
\end{align}
To proceed, we first note that ${Q(n|{\bf a})}/{ {Q(N|{\bf a})}^{n/N} }=1$ for $n=0$ and $n=N$. Considering the upper limit of the first term in Eq.\,\ref{eq:int3}
\begin{align}
\underset{\theta \rightarrow \infty}  \lim \theta \frac {\sum_n n {\rm e}^{\theta n} \frac{Q(n|{\bf a})}{ {Q(N|{\bf a})}^{n/N} }}{ {\sum_n {\rm e}^{\theta n} \frac{Q(n|{\bf a})}{ {Q(N|{\bf a})}^{n/N} }}}
=
\underset{\theta \rightarrow \infty}  \lim \theta \frac{N + (N-1){\rm e}^{-\theta} \frac{Q(N-1|{\bf a})}{ {Q(N|{\bf a})}^{(N-1)/N )}}+O({\rm e}^{-2\theta})}{1+{\rm e}^{-\theta} \frac{Q(N-1|{\bf a})}{ {Q(N|{\bf a})}^{(N-1)/N )}}+O({\rm e}^{-2\theta})} =   N \theta  .
\end{align}
Similarly, the lower limit of the first term of Eq.\,\ref{eq:int3} is 0, since the all terms are exponentially suppressed relative to $n=0$ as $\theta \rightarrow -\infty$. Turning to the upper limit of the second term in Eq.\,\ref{eq:int3},
\begin{align}
\underset{\theta \rightarrow \infty}  \lim \ln {\sum_n {\rm e}^{\theta n} \frac{Q(n|{\bf a})}{ {Q(N|{\bf a})}^{n/N} }}
=
\underset{\theta \rightarrow \infty}  \lim  N\theta +  \ln \left(1 + {\rm e}^{-\theta} \frac{Q(N-1|{\bf a})}{ {Q(N|{\bf a})}^{(N-1)/N )}}+O({\rm e}^{-2\theta}) \right)=   N\theta.
\end{align}
Similarly,  the lower limit of the second term of Eq.\,\ref{eq:int3} is 0, since the only term not exponentially suppressed is $\ln 1$ rather than $\ln {\rm e}^{N\theta}$. Combining all contributions shows that the integral in Eq.\,\ref{eq:int3} is identically zero. Thus 
\begin{equation}
\langle w_{\rm pol}({\bf a}) \rangle =- k_{\rm B}T \ln Q(N|{\bf a}), 
\end{equation}
as required.

\section{S2. Evaluation of $\langle w^f \rangle $} 
To calculate the total work for copying a given ${\bf a}$, we sum ${\langle w_{\rm pol}({\bf a})\rangle }$ with $\langle w_{\rm sep}({\bf a})\rangle =  -\Delta F_{\rm s} - C - \Delta F_{AB}({\bf a}) $ and $\langle w_{\rm seed} \rangle = \Delta F_{\rm s} + C$, finding
\begin{equation}
\langle w ({\bf a})\rangle =  -{k_{\rm B}T} {\ln Q(N|{\bf a})} -\Delta F_{AB}({\bf a}).
\end{equation}
Since $Q(N|{\bf a})$ is a partition function,
\begin{equation}
p_f({\bf b}|{\bf a}) = \frac{ \prod_{n=1}^N{\rm e}^{-\beta \Delta F_{b_n}} ((1-\delta_{a_n, b_n}){\rm e}^{-\beta \Delta F_{\rm nc}} + \delta_{a_n, b_n}{\rm e}^{-\beta \Delta F_{\rm c}})}{Q(N|{\bf a}) }.
\end{equation}
Thus, taking the definition of ${\Delta F_{AB}}({\bf a})$ from the main text, 
\begin{equation}
{\Delta F_{AB}}({\bf a})=   k_{\rm B} T\sum_{{\bf b}} p_f({\bf b}|{\bf a}) \ln \left( p_f({\bf b}|{\bf a}) Q(N|{\bf a)} \prod_{n=1}^{N} {\rm e}^{\beta \Delta F_{b_n}}   \right),
\end{equation} 
and hence
\begin{equation}
\langle w({\bf a})\rangle =  k_{\rm B} T  \sum_{{\bf b}} p_f({\bf b}|{\bf a}) \sum_{n=1}^N  \Delta F_{b_n} -T \mathcal{H}[p_f({\bf b}|{\bf a})].
\end{equation}
Averaging  over $p({\bf a})$ and using the fact that $|{\bf b}|$ is guaranteed to be eqaul to $|{\bf a}|=N$ at the end of the protocol outlined, along with
 $\mathcal{H}(B|A) = \mathcal{H}(B) - k_{\rm B}T \mathcal{I}(A;B)$, gives the desired result in Eq. 5 of the main text.

\section{S3. Non-random copying of multiple templates}
\label{sec:non-random}
Eq. 7 of the main text the was derived assuming that each of the $M$ copies was based on a template ${\bf a}$ with a probability $p({\bf a})$. Thus the total number of copies of each template is uncertain. An alternative protocol might make guarantee to make $M^{\bf a}$ copies of template ${\bf a}$, with the only uncertainty coming from finite accuracy ($p_f({\bf b}| {\bf a})$ has non-zero entropy). 

Assume for simplicity that for each ${\bf a}$, $p_f({\bf b}| {\bf a})$ is only non-zero for at most a single ${\bf a}$ for a given ${\bf b}$. In this limit, copies of each template are perfectly distinguishable, even though they are not deterministic. In this case, the total free energy is simply the sum of the free energies of the copies of each template, which follows from Eq. 7 of the main text as
\begin{align}
&\mathscr{F}[\phi(\{M_{\bf b}\})] = -k_{\rm B} T \sum_{\bf a}  \left(M^{\bf a} \sum_{\bf b} p_f({\bf b}| {\bf a}) \ln  Z_{\bf b} -  M^{\bf a}\sum_{\bf b} p_f({\bf b} |{\bf a}) \ln  p_f({\bf b} |{\bf a}) -  \ln  M^{\bf a}! \right).
\label{eq:F_nophi}
\end{align}
We define $\psi({\bf a}) = M^{\bf a}/M$, and  $\psi({\bf b}) = \sum_{\bf a} \psi({\bf a}) p_f({\bf b} | {\bf a})$: 
\begin{align}
&\mathscr{F}[\phi(\{M_{\bf b}\})] = -k_{\rm B} T M \sum_{\bf b} \psi({\bf b}) \ln Z_{\bf b} + k_{\rm B} T M \sum_{\bf b} \psi({\bf b}) \ln \psi({\bf b}) +  k_{\rm B} T \sum_{\bf  a} M^{\bf a} \ln \frac{M}{M^{\bf a}} + k_{\rm B} T \sum_{\bf a}  \ln M^{\bf a}!.
\end{align}
The above result uses the fact that, if $p_f({\bf b}| {\bf a})$ is only non-zero for at most a single ${\bf b}$, $\sum_{\bf a} \psi({\bf a}) p_f({\bf b} | {\bf a}) \ln p_f({\bf b} | {\bf a}) =  \sum_{\bf b} \psi({\bf b}) \ln \frac{ \psi({\bf b})}{\psi({\bf a})}$. Simplifying further,
\begin{align}
&\mathscr{F}[\phi(\{M_{\bf b}\})] = -k_{\rm B} T M \sum_{\bf b} \psi({\bf b}) \ln Z_{\bf b} + k_{\rm B} T M \sum_{\bf b} \psi({\bf b}) \ln \psi({\bf b}) + k_{\rm B} T M \ln M - k_{\rm B} T \sum_{\bf  a} M^{\bf a} \ln {M^{\bf a}} + k_{\rm B} T \sum_{\bf a}  \ln M^{\bf a}!.
\label{eq:check1}
\end{align}
Comparing to Eq. 7 of the main text, we see that the first two terms are directly equivalent if we take $\psi({\bf a}) = p({\bf a})$, {\it ie.,} map the (deterministic) fraction of polymers that are copies of ${\bf a}$ to the probability of copying ${\bf a}$ in the original context. The remaining terms, however, are not identical. This is because, although the average number of copies of any template ${\bf a}$ is correctly estimated using this mapping, there is additional entropy in the system described by Eq. 7 of the main text since the number of copies of ${\bf a}$  fluctuates around $Mp({\bf a})$, whereas in the system described by Eq. \,\ref{eq:check1}, there are always $M^{\bf a} = M \psi({\bf a})$ copies of ${\bf a}$. 

In the limit of large $M^{\bf a}$, these fluctuations are relatively small. In this case, $\sum_{\bf a}  \ln M^{\bf a}! \approx  \sum_{\bf a} M^{\bf a} \ln M^{\bf a} - M$, and $ M \ln M - M \approx \ln M!$. Thus,
\begin{align}
&\mathscr{F}[\phi(\{M_{\bf b}\})] \approx -k_{\rm B} T M \sum_{\bf b} \psi({\bf b}) \ln Z_{\bf b} + k_{\rm B} T M \sum_{\bf b} \psi({\bf b}) \ln \psi({\bf b}) + k_{\rm B} T \ln M!,
\label{eq:check2}
\end{align}
and the stored free energy is essentially equal to that of a system in which copies of template ${\bf a}$ are made randomly with probability $p({\bf a}) = \psi({\bf a}) = M^{\bf a}/M$, resulting in an output distribution $p_f({\bf b})$ of each copy (Eq. 7 of the main text).

\section{S4. A protocol for efficient copying of multiple templates}
We will work within the seed-assisted polymerization model analysed in the main text, and again consider the case in which there is no overlap between the probability distribution of copies $p_f({\bf b}| {\bf a})$ for distinct ${\bf a}$ sequences. Consider the protocol illustrated in Fig.\,\ref{fig:protocol_multi}. Initially, we start with $M$ seeds in the large volume. We then reversibly transfer each of these seeds to a number of smaller volumes that each contain a known polymer of type $A$, using a biochemical ``hook'' that can bind to the seeds. It must be possible to quasistatically increase the strength with which this hook binds to the seeds, for example by varying the solution conditions, to make the pick up/deposit efficient. Such a system may be challenging to engineer, but does not violate the laws of thermodynamics. Once inside the small volumes, a copy of the relevant $A$ polymer is grown from each of the seeds using the protocol outlined in the main text. The seeds can then be returned to the large volume using the biochemical hooks. 

\begin{figure}
\centering
\includegraphics[width=170mm]{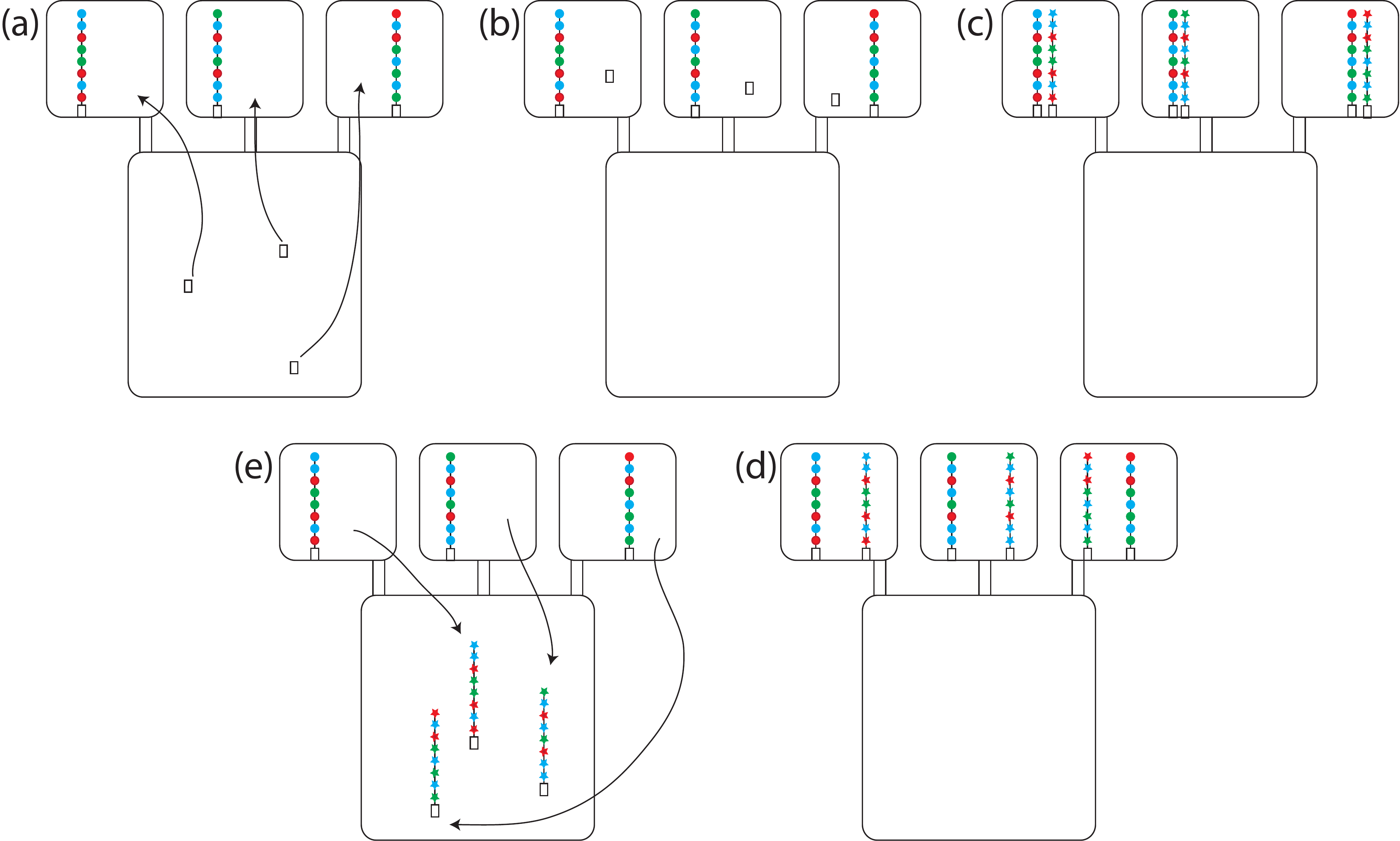}
\caption{Protocol for producing copies of multiple templates. (a) Initially, seeds are present in a large volume, before being transferred to smaller volumes, each containing a template. In steps (b)-(d), the seeds are brought into contact with the templates, polymerization is driven by adjusting the chemical potential of monomers, and copied polymers are separated from their templates, as outlined in more detail in the main text. (e) Copied polymers are returned to the large volume, either using the same biochemical `hooks' as in step (a), or hooks that are specific to  the known template sequence in each small volume. 
  \label{fig:protocol_multi}}
\end{figure}

First, let us identify the free energy change due to the operation. Following Eq. 6 of the main text, the initial state of $M$ seeds has free energy $\mathcal{F}_i = -k_{\rm B}T M \ln Z_0  + k_{\rm B}T M!$. The free energy of the final state is given by Eq.\,\ref{eq:F_nophi}, since the number of copies of each template sequence if known. Thus
\begin{align}
\Delta \mathscr{F} = -k_{\rm B} T \sum_{\bf a}  \left(M^{\bf a} \sum_{\bf b} p_f({\bf b}| {\bf a}) \ln \left( \frac{Z_{\bf b}}{Z_0} \right) -  M^{\bf a}\sum_{\bf b} p_f({\bf b} |{\bf a}) \ln  p_f({\bf b} |{\bf a}) -  \ln  M^{\bf a}! \right) -  k_{\rm B}T \ln M!.
\label{eq:dFmultimix}
\end{align} 
This free energy change of course sets the minimum work necessary to complete the operation. Does the proposed protocol achieve it? The cost of the protocol for steps (b) to (d) in Fig.\,\ref{fig:protocol_multi} follow from the calculation for a single copy in the main text; we simply need to sum over all ${\bf a}$ sequences.  Since the hook binds only to the seeds, the transfer processes ((a) and (e) in Fig.\,\ref{fig:protocol_multi}) are effectively inverse operations on the seeds and the work done during the transfer processes cancels. Thus, proceeding as with Eq. 5 of the main text and using  $-k_{\rm B}T \ln (Z_{\rm b}/Z_0) =    \sum_{n=1}^{|{\bf b}|}  \Delta F_{b_n}$, 
\begin{align}
\langle W \rangle =  -k_{\rm B} T \sum_{\bf a}  \left(M^{\bf a} \sum_{\bf b} p_f({\bf b}| {\bf a}) \ln \left( \frac{Z_{\bf b}}{Z_0} \right) -  M^{\bf a}\sum_{\bf b} p_f({\bf b} |{\bf a}) \ln  p_f({\bf b} |{\bf a})\right).
\label{eq:w1multimix}
\end{align} 
We immediately see that $\langle W \rangle - \Delta \mathscr{F} = k_{\rm B}T M! - k_{\rm B}T \sum_{\bf a} \ln  M^{\bf a}! >0 $ (assuming more than one sequence ${\bf a}$ is copied). The protocol proposed is   therefore irreversible. The fundamental reason is that, in returning the copied sequences to the large volume, distinct molecules are allowed to mix irreversibly  (work is not extracted from this mixing). Since the hook only binds to seeds and cannot distinguish between polymers, transferring seeds in and out of the large volume appear to be inverse processes, whereas in fact they are not. 

As in Section S3, we can introduce $\phi({\bf a}) = M^{\bf a}/M$ and $\phi({\bf b}) = \sum_{\bf a} \phi({\bf a}) p_f({\bf b}| {\bf a})$. Again, in the limit of large $M$, the error associated with interpreting $\phi({\bf a})$ as a probability of copying ${\bf a}$ becomes relatively small, and in this case Eq.\,\ref{eq:dFmultimix} (following Section S3) can be interpreted as 
\begin{align}
\Delta \mathscr{F} \approx -k_{\rm B} T M \sum_{\bf b} \phi({\bf b})  \ln \left( \frac{Z_{\bf b}}{Z_0} \right) -k_{\rm B} T M \mathcal{H}(B),
\end{align}
and Eq.\,\ref{eq:w1multimix} as
\begin{align}
\langle W \rangle \approx -k_{\rm B} T M \sum_{\bf b} \phi({\bf b})  \ln \left( \frac{Z_{\bf b}}{Z_0} \right) -k_{\rm B} T M \mathcal{H}(B|A).
\end{align} 
These results imply a dissipated work per polymer $T \Delta \sigma =\langle W \rangle - \Delta \mathscr{F} \approx k_{\rm B}T \mathcal{I}(A;B)$, consistent with the observation in the main text for a system in which templates are chosen in a genuinely random fashion, and mixing is irreversible.  Thus, if mixing occurs irreversibly, the entropy of the universe increases by  $T \Delta \sigma \approx k_{\rm B}T \mathcal{I}(A;B)$

An alternative approach would be to return seeds to the large volume using a range of biochemical hooks that are 100\% selective for the products of each template sequence. Again, such a system may be difficult to engineer, but is not physically impossible. In this case, more work is extracted upon returning the polymers to the large volume than was required to transfer the seeds out originally, because  it is easier to systematically  release a molecule into solution using a selective hook that can only bind to a subset of the molecules present rather than a generic hook that will bind to any of them. Consider, for example, releasing a polymer into a pool of $L$ polymers that can all bind to the hook with the same affinity. An efficient protocol would involve slowly adjusting conditions so that the binding free energy of a single polymer, $\Delta F_h$, goes from from $-\infty$ to $+\infty$.  During this process, the probability that any polymer is bound to this non-specific hook is given by 
\begin{equation}
p_{\rm non-spec}(\Delta F_h) = \frac{L \exp(-\Delta F_h/k_{\rm B}T)}{1+ L \exp(-\Delta F_h/k_{\rm B}T)}   
\end{equation}
For a specific hook that only binds to $L_{\bf a}<L$ polymers with the same affinity, 
\begin{equation}
p_{\rm spec}(\Delta F_h) = \frac{L_{\bf a} \exp(-\Delta F_h/k_{\rm B}T)}{1+ L_{\bf a} \exp(-\Delta F_h/k_{\rm B}T)}   
\end{equation}
Since $p_{\rm spec}(\Delta F_h) < p_{\rm non-spec}(\Delta F_h)$, $\Delta F_h$ will need to be raised less far before the specific hook is typically free of polymers, implying that less work must be done. Specifically,
\begin{equation}
\langle w_{\rm non-spec} \rangle - \langle w_{\rm spec} \rangle = \int_{-\infty}^{+\infty} {\rm d} \Delta  F_h  \, \left( p_{\rm non-spec}(\Delta F_h) - p_{\rm spec}(\Delta F_h) \right)= k_{\rm B} T \ln \left(\frac{L}{L_{\bf a}} \right).
\end{equation}
Summing this difference over all added polymers (and remembering that the number of polymers in the pool increases as more are returned) gives a reduction in cost due to specificity of $k_{\rm B}T\ln M! - k_{\rm B} T \sum_{\bf a} \ln M^{\bf a}!$. This result could have been anticipated by noting that the specific hooks do the work required to create a solution of $M_{\bf a}$ polymers for each ${\bf a}$,  whereas the non-specific hooks do the work required to create a solution of $M$ polymers. Augmenting Eq.\,\ref{eq:w1multimix} yields
\begin{align}
\langle W_{\rm selective} \rangle =  -k_{\rm B} T \sum_{\bf a}  \left(M^{\bf a} \sum_{\bf b} p_f({\bf b}| {\bf a}) \ln \left( \frac{Z_{\bf b}}{Z_0} \right) -  M^{\bf a}\sum_{\bf b} p_f({\bf b} |{\bf a}) \ln  p_f({\bf b} |{\bf a})-  \ln  M^{\bf a}! \right) -  k_{\rm B}T \ln M! = \Delta \mathscr{F},
\end{align} 
indicating that this selective protocol is reversible. Indeed, reversing the procedure constitutes measuring the sequences and depolymerizing using the appropriate template, the necessary procedure for thermodynamically efficient depolymerization identified in the main text. With such a template-specific depolymerization protocol, the net work over the full cycle of polymerisation and depolymerisation is zero, reflecting that the cycle is reversible. However, as discussed in the main text, and addressed in the next SI section, inside cells, depolymerization occurs in a generic, non-template-specific fashion, in which case the depolymerization process (and hence the full cycle of polymerization and depolymerization) is necessarily irreversible.

\section{S5. Evaluation of work during depolymerization}
Our  first goal is to evaluate
\begin{align}
\langle W^z_{\rm depol} \rangle =- M \int_{-\infty}^{\bar{\mu}_z} {\rm d} \mu \, \mu \frac{{\rm d} \langle|{\bf b}| \rangle}{{\rm d}\mu}, 
\end{align}
in which the manipulation of chemical potential is quasistatic so that  $\langle|{\bf b}| \rangle$ is determined by the equilibrium distribution at any given $\mu$, and $\langle|{\bf b}| \rangle = \bar{N}_{z} $ at $\mu = \bar{\mu}_z$. Further, for isolated $B$ polymers,
\begin{align}
\frac{P(|{\bf b}|=n)}{P(|{\bf b}|=0)} = {\rm e}^{{ \mu n}/{k_{\rm B}T}} \left(\sum_{x=1}^m {\rm e}^{- \Delta F_x/k_{\rm B}T} \right)^n ={\rm e}^{{ \mu n}/{k_{\rm B}T}} \omega^n = {\rm e}^{ \theta n}.
\label{eq:relative_P}
\end{align}
The above equation defines $\omega$ and $\theta = \mu/k_{\rm B}T + \ln \omega$. Thus
\begin{align}
\langle W^z_{\rm depol} \rangle =- k_{\rm B}T M \int_{-\infty}^{\bar{\theta}_z} {\rm d} \theta \,  (\theta - \ln \omega) \frac{{\rm d}\langle|{\bf b}| \rangle }{{\rm d} \theta}. 
\label{eq:integral1}
\end{align}
The second term can be evaluated directly,
\begin{align}
\int_{-\infty}^{\bar{\theta}_z} {\rm d} \theta \,   \ln \omega \frac{{\rm d}\langle|{\bf b}| \rangle }{{\rm d} \theta} = \ln \omega \left[ \langle|{\bf b}| \rangle \right]_{-\infty}^{\bar{\theta}_z} = \bar{N}_{z} \ln \omega,
\label{eq:term1}
\end{align}
since the upper limit of the integral is such that $ \langle|{\bf b}| \rangle = \bar{N}_{z}$ by design, and $ \langle|{\bf b}| \rangle =0$ at the lower limit. For the second term, we use the fact that 
\begin{align}
\langle|{\bf b}| \rangle = \frac{\rm d}{{\rm d} \theta} \ln \sum_{|{\bf b}| =0}^\infty {\rm e}^{\theta n} = -\frac{\rm d}{{\rm d} \theta}  \ln(1- {\rm e}^\theta) = \frac{{\rm e}^{\theta}}{1-{\rm e}^\theta}.
\label{eq:<b>}
\end{align}
Thus
\begin{align}
 \int_{-\infty}^{\bar{\theta}_z} {\rm d} \theta \,  \theta  \frac{{\rm d}\langle|{\bf b}| \rangle }{{\rm d} \theta} =  \left[ \theta \langle|{\bf b}| \rangle \right]_{-\infty}^{\bar{\theta}_z} +  \int_{-\infty}^{\bar{\theta}_z} {\rm d} \theta \, \frac{\rm d}{{\rm d} \theta}  \ln(1- {\rm e}^\theta).
\end{align}
Evaluating,
\begin{align}
 \int_{-\infty}^{\bar{\theta}_z} {\rm d} \theta \,  \theta  \frac{{\rm d}\langle|{\bf b}| \rangle }{{\rm d} \theta} =  \left(\frac{\bar{\mu}_z}{k_{\rm B}T} + \ln \omega \right) \bar{N}_{z} +   \ln \left(1- \omega \exp \left(\frac{\bar{\mu}_{z} }{k_{\rm B}T} \right)\right).
\label{eq:term2}
\end{align}
Combining Eq.\,\ref{eq:term1}, \ref{eq:term2} and \ref{eq:integral1}, we find
\begin{align} 
\langle W^z_{\rm depol} \rangle = -k_{\rm B} T M \left(\frac{\bar{\mu}_{z}}{k_{\rm B}T}  \bar{N}_{z} +   \ln \left(1- \omega \exp \left(\frac{\bar{\mu}_{}}{k_{\rm B}T} \right)\right)  \right).
\end{align}
To further simplify, we note that since $\bar{N}_{z}  = {\rm e}^{ \bar{\theta}_{z} } / (1-{\rm e}^{ \bar{\theta}_{z} })$ from Eq.\,\ref{eq:<b>}, $\bar{\theta}_{z}  = \bar{\mu}_{z} /k_{\rm B}T + \ln \omega = \ln \left(\bar{N}_{z}  /(1+\bar{N}_{z}  ) \right)$. Thus
\begin{align} 
\langle W^z_{\rm depol} \rangle = k_{\rm B} T M \bar{N}_{z}  \ln \omega - k_{\rm B} T M \bar{N}_{z}  \ln \bar{N}_{z}  +  k_{\rm B} T M (\bar{N}_{z} +1) \ln  (\bar{N}_{z} +1) .
\label{eq:wdepol_result1}
\end{align}

We will now show that $\langle W^z \rangle_{\rm rev} - \langle W^z_{\rm depol} \rangle$  is identical to Eq.  of the main text, verifying that the template-free non-specific depolmerization protocol leads to the expected dissipation for this model.
 From Eq. 7 of the main text, it follows by definition  that 
\begin{align}
\langle W^z \rangle_{\rm rev} = - k_{\rm B} T M \sum_{\bf b} p_z({\bf b})  \ln  \frac{Z_{\bf b}} {Z_0} 
  - k_{\rm B} T M \mathcal{H}[p_z({\bf b})],
\end{align}
which is the difference in free energy between the distribution of macrostates $\phi(\{M_{\bf b}\}) = M!\prod_{\bf b}  p_z({\bf b})^{M_{\bf b}}/M_{\bf b}!$ and the template-only macrostate. It thus remains to show that  our protocol of depolymerization recovers exactly the difference between the free energy stored in the equilibrium distribution of average length $\bar{N}_z$ and the seed-only state:
\begin{align}
\langle W^z_{\rm depol} \rangle =  - k_{\rm B} T M \sum_{\bf b} p^{\rm eq}_{\bar{N}_z}({\bf b})  \ln  \frac{Z_{\bf b}} {Z_0} 
  - k_{\rm B} T M \mathcal{H}[ p^{\rm eq}_{\bar{N}_z}({\bf b})],
\label{eq:proof1}
\end{align}
For the model in question, the equilibrium distribution of sequences at $\bar{\mu}_{z} $ is 
\begin{align}
p^{\rm eq}_{\bar{N}_{z} }({\bf b}) = \frac{ {\rm e}^{\bar{\mu}_{z}  |{\bf b}|/k_{\rm B}T} Z_0 \prod_{x=1}^{|{\bf b}|} {\rm e}^{- \Delta F_{b_x}/k_{\rm B}T}} {\Omega} = \frac{ {\rm e}^{\bar{\mu}_{z}  |{\bf b}|/k_{\rm B}T}  Z_{\bf b}}{\Omega},
\end{align}
in which $\Omega = \sum_{\bf b}  {\rm e}^{\bar{\mu}_{z}  |{\bf b}|/k_{\rm B}T} Z_0 \prod_{x=1}^{|{\bf b}|} {\rm e}^{- \Delta F_{b_x}/k_{\rm B}T} $ is a normalizing partition function. Substituting into the RHS of Eq.\,\ref{eq:proof1}, we  obtain
\begin{align}
 k_{\rm B} T M \sum_{\bf b} p^{\rm eq}_{\bar{N}_{z} }({\bf b})  \ln \frac{\Omega}{Z_0} -  k_{\rm B} T M \sum_{\bf b} p^{\rm eq}_{\bar{N}_{z} }({\bf b}) \frac{\bar{\mu}_{z}  }{k_{\rm B} T} |{\bf b}|,
\end{align}
which simplifies to
\begin{align}
k_{\rm B} T M  \ln \left( \sum_{\bf b}  {\rm e}^{\bar{\mu}_{z}  |{\bf b}|/k_{\rm B}T}\prod_{x=1}^{|{\bf b}|} {\rm e}^{- \Delta F_{b_x}/k_{\rm B}T} \right)-  k_{\rm B} T M \bar{N}_{z}  \frac{\bar{\mu}_{z}  }{k_{\rm B} T}.
\end{align}
Since all terms in the sum with the same $|{\bf b}|$ have the same prefactor, and re-using the original definiton of $\theta$ in Eq.\,\ref{eq:relative_P}, we can rewrite the RHS of Eq.\,\ref{eq:proof1} as 
\begin{align}
 k_{\rm B} T M  \ln \left( \sum_{|{\bf b}|}  {\rm e}^{\bar{\theta}_{z}  |{\bf b}|} \right)-  k_{\rm B} T M \bar{N}_{z}  \frac{\bar{\mu}_{z}  }{k_{\rm B} T} =  -k_{\rm B} T M \ln\left(1-{\rm e}^{\bar{\theta}_{z} } \right)- k_{\rm B} T M \bar{N}_{z}  \frac{\bar{\mu}_{z}  }{k_{\rm B} T} .
\end{align}
Using  $-\ln\left(1-{\rm e}^{\bar{\theta}_{z} }\right) = \ln(1+\bar{N}_{z} )$ and $\frac{\bar{\mu}_{z}  }{k_{\rm B} T} = -\ln \omega +  \ln \left(\bar{N}_{z}  /(1+\bar{N}_{z}  )\right)$, as justified above,  the RHS of Eq.\,\ref{eq:proof1} becomes
\begin{align}
 k_{\rm B} T M \ln(1+\bar{N}_{z} ) + k_{\rm B} T M \bar{N}_{z}  \ln \omega -  k_{\rm B} T M \bar{N}_{z}  \ln \left(\bar{N}_{z}  )/(1+\bar{N}_{z} ) \right).
\end{align}
This expression is trivially equal to $\langle W^z_{\rm depol} \rangle $  as expressed in Eq.\,\ref{eq:wdepol_result1}, confirming our claim that this protocol recovers only the work stored in the equilibrium state of average length $\bar{N}_z$, and hence that the overall entropy generated during reversible polymerization followed by non-selective (irreversible) depolymerization is
\begin{equation}
T\Delta \sigma = \langle W^z \rangle_{\rm rev} - \langle W^z_{\rm depol} \rangle =  - k_{\rm B} T M \sum_{\bf b} \left(p_z({\bf b}) -  p^{\rm eq}_{\bar{N}_{z} }({\bf b}) \right) \ln {Z_{\bf b}}
  - k_{\rm B} T M \mathcal{H}[p_z({\bf b})]+ k_{\rm B} T M \mathcal{H}[ p^{\rm eq}_{\bar{N}_{z} }({\bf b})], 
\end{equation} 
which for this model is 
\begin{align}
T\Delta \sigma = \langle W^z \rangle_{\rm rev} - \langle W^z_{\rm depol} \rangle =   M \sum_{\bf b} p_z({\bf b}) \sum_{x=1}^{|{\bf b}|} \Delta F_{b_x}
  - k_{\rm B} T M \mathcal{H}[p_z({\bf b})]\\
 - \left(  k_{\rm B} T M \bar{N}_{z}  \ln\left(\sum_{x=1}^m {\rm e}^{- \Delta F_x/k_{\rm B}T}   \right) - k_{\rm B} T M \bar{N}_{z}  \ln \bar{N}_{z}  +  k_{\rm B} T M (\bar{N}_{z} +1) \ln  (\bar{N}_{z} +1) \right). \nonumber
\end{align} 

\section{S6. Accurate and reversible production of persistent copies in non-autonomous systems}
\teob{
In the main text, we argue that in an autonomous, continuously-operating system producing persistent copies, the template  can only act as a catalyst. Specificity of copy sequences can only follow from stabilization of intermediates and hence copy-template interactions can only provide a kinetic, rather than overall thermodynamic, discrimination. Kinetic discrimination only functions out of equilibrium, and hence we argue that unlike in templated self-assembly, autonomous production of accurate persistent copies requires dissipation (entropy generation) for finite accuracy.}

\teob{However, we also discuss a protocol for reversible production of
  persistent copies in which a template is used to produce a
  sequence-specific copy without an overall increase in the entropy of
  the universe. This is possible because the system is not autonomous,
 operating continuously under fixed external
  conditions. Instead, an experimenter varies the conditions
  periodically, allowing reversible self-assembly to be subsequently
  followed by separation. The key point is that through a
  time-dependent control mechanism, a system can be driven through a
  series of states: attach seed; grow; detach, without
  dissipating. This fact enables the sequence-specific copy-template
  interactions that favor growth of specific B sequences whilst in
  contact with A to be manifest in the final sequence, since
  detachment occurs at the desired time regardless of the copied
  sequence. In an autonomous, quasi-reversible setting, the tendency
  of accurate sequences to to stick to the template will favor the
  attachment of certain monomers, but will interfere equally with the
  subsequent detachment. Of course, our statement that the driven
  system involves no entropy production neglects any additional costs
  inherent to implementing the experimenter's control protocol. }
\end{widetext}

\begin{thebibliography}{30}%
\makeatletter
\providecommand \@ifxundefined [1]{%
 \@ifx{#1\undefined}
}%
\providecommand \@ifnum [1]{%
 \ifnum #1\expandafter \@firstoftwo
 \else \expandafter \@secondoftwo
 \fi
}%
\providecommand \@ifx [1]{%
 \ifx #1\expandafter \@firstoftwo
 \else \expandafter \@secondoftwo
 \fi
}%
\providecommand \natexlab [1]{#1}%
\providecommand \enquote  [1]{``#1''}%
\providecommand \bibnamefont  [1]{#1}%
\providecommand \bibfnamefont [1]{#1}%
\providecommand \citenamefont [1]{#1}%
\providecommand \href@noop [0]{\@secondoftwo}%
\providecommand \href [0]{\begingroup \@sanitize@url \@href}%
\providecommand \@href[1]{\@@startlink{#1}\@@href}%
\providecommand \@@href[1]{\endgroup#1\@@endlink}%
\providecommand \@sanitize@url [0]{\catcode `\\12\catcode `\$12\catcode
  `\&12\catcode `\#12\catcode `\^12\catcode `\_12\catcode `\%12\relax}%
\providecommand \@@startlink[1]{}%
\providecommand \@@endlink[0]{}%
\providecommand \url  [0]{\begingroup\@sanitize@url \@url }%
\providecommand \@url [1]{\endgroup\@href {#1}{\urlprefix }}%
\providecommand \urlprefix  [0]{URL }%
\providecommand \Eprint [0]{\href }%
\providecommand \doibase [0]{http://dx.doi.org/}%
\providecommand \selectlanguage [0]{\@gobble}%
\providecommand \bibinfo  [0]{\@secondoftwo}%
\providecommand \bibfield  [0]{\@secondoftwo}%
\providecommand \translation [1]{[#1]}%
\providecommand \BibitemOpen [0]{}%
\providecommand \bibitemStop [0]{}%
\providecommand \bibitemNoStop [0]{.\EOS\space}%
\providecommand \EOS [0]{\spacefactor3000\relax}%
\providecommand \BibitemShut  [1]{\csname bibitem#1\endcsname}%
\let\auto@bib@innerbib\@empty
\bibitem [{\citenamefont {Alberts}\ \emph {et~al.}(2002)\citenamefont
  {Alberts}, \citenamefont {Johnson}, \citenamefont {Lewis}, \citenamefont
  {Raff}, \citenamefont {Roberts},\ and\ \citenamefont {Walter}}]{Alberts2002}%
  \BibitemOpen
  \bibfield  {author} {\bibinfo {author} {\bibfnamefont {B.}~\bibnamefont
  {Alberts}}, \bibinfo {author} {\bibfnamefont {A.}~\bibnamefont {Johnson}},
  \bibinfo {author} {\bibfnamefont {J.}~\bibnamefont {Lewis}}, \bibinfo
  {author} {\bibfnamefont {M.}~\bibnamefont {Raff}}, \bibinfo {author}
  {\bibfnamefont {K.}~\bibnamefont {Roberts}}, \ and\ \bibinfo {author}
  {\bibfnamefont {P.}~\bibnamefont {Walter}},\ }\href@noop {} {\emph {\bibinfo
  {title} {Molecular Biology of the Cell, 4th Edition}}}\ (\bibinfo
  {publisher} {Garland Science, New York},\ \bibinfo {year} {2002})\BibitemShut
  {NoStop}%
\bibitem [{\citenamefont {Bennett}(1979)}]{Bennett1979}%
  \BibitemOpen
  \bibfield  {author} {\bibinfo {author} {\bibfnamefont {C.~H.}\ \bibnamefont
  {Bennett}},\ }\href@noop {} {\bibfield  {journal} {\bibinfo  {journal}
  {Biosystems}\ }\textbf {\bibinfo {volume} {11}},\ \bibinfo {pages} {85}
  (\bibinfo {year} {1979})}\BibitemShut {NoStop}%
\bibitem [{\citenamefont {Cady}\ and\ \citenamefont {Qian}(2009)}]{Cady2009}%
  \BibitemOpen
  \bibfield  {author} {\bibinfo {author} {\bibfnamefont {F.}~\bibnamefont
  {Cady}}\ and\ \bibinfo {author} {\bibfnamefont {H.}~\bibnamefont {Qian}},\
  }\href {http://stacks.iop.org/1478-3975/6/i=3/a=036011} {\bibfield  {journal}
  {\bibinfo  {journal} {Phys. Biol.}\ }\textbf {\bibinfo {volume} {6}},\
  \bibinfo {pages} {036011} (\bibinfo {year} {2009})}\BibitemShut {NoStop}%
\bibitem [{\citenamefont {Andrieux}\ and\ \citenamefont
  {Gaspard}(2008)}]{Andrieux2008}%
  \BibitemOpen
  \bibfield  {author} {\bibinfo {author} {\bibfnamefont {D.}~\bibnamefont
  {Andrieux}}\ and\ \bibinfo {author} {\bibfnamefont {P.}~\bibnamefont
  {Gaspard}},\ }\href@noop {} {\bibfield  {journal} {\bibinfo  {journal} {Proc.
  Nat. Acad. Sci. USA}\ }\textbf {\bibinfo {volume} {105}},\ \bibinfo {pages}
  {9516} (\bibinfo {year} {2008})}\BibitemShut {NoStop}%
\bibitem [{\citenamefont {Sartori}\ and\ \citenamefont
  {Pigolotti}(2013)}]{Sartori2013}%
  \BibitemOpen
  \bibfield  {author} {\bibinfo {author} {\bibfnamefont {P.}~\bibnamefont
  {Sartori}}\ and\ \bibinfo {author} {\bibfnamefont {S.}~\bibnamefont
  {Pigolotti}},\ }\href {\doibase 10.1103/PhysRevLett.110.188101} {\bibfield
  {journal} {\bibinfo  {journal} {Phys. Rev. Lett.}\ }\textbf {\bibinfo
  {volume} {110}},\ \bibinfo {pages} {188101} (\bibinfo {year}
  {2013})}\BibitemShut {NoStop}%
\bibitem [{\citenamefont {Sartori}\ and\ \citenamefont
  {Pigolotti}(2015)}]{Sartori2015}%
  \BibitemOpen
  \bibfield  {author} {\bibinfo {author} {\bibfnamefont {P.}~\bibnamefont
  {Sartori}}\ and\ \bibinfo {author} {\bibfnamefont {S.}~\bibnamefont
  {Pigolotti}},\ }\href {\doibase 10.1103/PhysRevX.5.041039} {\bibfield
  {journal} {\bibinfo  {journal} {Phys. Rev. X}\ }\textbf {\bibinfo {volume}
  {5}},\ \bibinfo {pages} {041039} (\bibinfo {year} {2015})}\BibitemShut
  {NoStop}%
\bibitem [{\citenamefont {Luther}\ \emph {et~al.}(1998)\citenamefont {Luther},
  \citenamefont {Brandsch},\ and\ \citenamefont {{von
  Kiedrowski}}}]{Luther1998}%
  \BibitemOpen
  \bibfield  {author} {\bibinfo {author} {\bibfnamefont {A.}~\bibnamefont
  {Luther}}, \bibinfo {author} {\bibfnamefont {R.}~\bibnamefont {Brandsch}}, \
  and\ \bibinfo {author} {\bibfnamefont {G.}~\bibnamefont {{von Kiedrowski}}},\
  }\href@noop {} {\bibfield  {journal} {\bibinfo  {journal} {Nature}\ }\textbf
  {\bibinfo {volume} {396}},\ \bibinfo {pages} {245} (\bibinfo {year}
  {1998})}\BibitemShut {NoStop}%
\bibitem [{\citenamefont {Kim}\ \emph {et~al.}(2015)\citenamefont {Kim},
  \citenamefont {Lee}, \citenamefont {Hamada}, \citenamefont {Murata},\ and\
  \citenamefont {Park}}]{Kim2015}%
  \BibitemOpen
  \bibfield  {author} {\bibinfo {author} {\bibfnamefont {J.}~\bibnamefont
  {Kim}}, \bibinfo {author} {\bibfnamefont {J.}~\bibnamefont {Lee}}, \bibinfo
  {author} {\bibfnamefont {S.}~\bibnamefont {Hamada}}, \bibinfo {author}
  {\bibfnamefont {S.}~\bibnamefont {Murata}}, \ and\ \bibinfo {author}
  {\bibfnamefont {S.~H.}\ \bibnamefont {Park}},\ }\href@noop {} {\bibfield
  {journal} {\bibinfo  {journal} {Nat. Nanotechnol.}\ }\textbf {\bibinfo
  {volume} {10}},\ \bibinfo {pages} {528} (\bibinfo {year} {2015})}\BibitemShut
  {NoStop}%
\bibitem [{\citenamefont {Sadownik}\ \emph {et~al.}(2016)\citenamefont
  {Sadownik}, \citenamefont {Mattia}, \citenamefont {Nowak},\ and\
  \citenamefont {Otto}}]{Sadownik2016}%
  \BibitemOpen
  \bibfield  {author} {\bibinfo {author} {\bibfnamefont {J.~W.}\ \bibnamefont
  {Sadownik}}, \bibinfo {author} {\bibfnamefont {E.}~\bibnamefont {Mattia}},
  \bibinfo {author} {\bibfnamefont {P.}~\bibnamefont {Nowak}}, \ and\ \bibinfo
  {author} {\bibfnamefont {S.}~\bibnamefont {Otto}},\ }\href@noop {} {\bibfield
   {journal} {\bibinfo  {journal} {Nat. Chem.}\ }\textbf {\bibinfo {volume}
  {8}},\ \bibinfo {pages} {264} (\bibinfo {year} {2016})}\BibitemShut {NoStop}%
\bibitem [{\citenamefont {Schulman}\ \emph {et~al.}(2012)\citenamefont
  {Schulman}, \citenamefont {Yurke},\ and\ \citenamefont
  {Winfree}}]{Schulman2012}%
  \BibitemOpen
  \bibfield  {author} {\bibinfo {author} {\bibfnamefont {R.}~\bibnamefont
  {Schulman}}, \bibinfo {author} {\bibfnamefont {B.}~\bibnamefont {Yurke}}, \
  and\ \bibinfo {author} {\bibfnamefont {E.}~\bibnamefont {Winfree}},\
  }\href@noop {} {\bibfield  {journal} {\bibinfo  {journal} {Proc. Nat. Acad.
  Sci. USA}\ }\textbf {\bibinfo {volume} {109}},\ \bibinfo {pages} {6405}
  (\bibinfo {year} {2012})}\BibitemShut {NoStop}%
\bibitem [{\citenamefont {Orgel}(1992)}]{Orgel1992}%
  \BibitemOpen
  \bibfield  {author} {\bibinfo {author} {\bibfnamefont {L.~E.}\ \bibnamefont
  {Orgel}},\ }\href@noop {} {\bibfield  {journal} {\bibinfo  {journal}
  {Nature}\ }\textbf {\bibinfo {volume} {358}},\ \bibinfo {pages} {203}
  (\bibinfo {year} {1992})}\BibitemShut {NoStop}%
\bibitem [{\citenamefont {Sievers}\ and\ \citenamefont {{von
  Kiedrowski}}(1994)}]{Sievers1994}%
  \BibitemOpen
  \bibfield  {author} {\bibinfo {author} {\bibfnamefont {D.}~\bibnamefont
  {Sievers}}\ and\ \bibinfo {author} {\bibfnamefont {G.}~\bibnamefont {{von
  Kiedrowski}}},\ }\href@noop {} {\bibfield  {journal} {\bibinfo  {journal}
  {Nature}\ }\textbf {\bibinfo {volume} {369}},\ \bibinfo {pages} {221}
  (\bibinfo {year} {1994})}\BibitemShut {NoStop}%
\bibitem [{\citenamefont {Vidonne}\ and\ \citenamefont
  {Philp}(2009)}]{Vidonne2009}%
  \BibitemOpen
  \bibfield  {author} {\bibinfo {author} {\bibfnamefont {A.}~\bibnamefont
  {Vidonne}}\ and\ \bibinfo {author} {\bibfnamefont {D.}~\bibnamefont
  {Philp}},\ }\href {\doibase 10.1002/ejoc.200800827} {\bibfield  {journal}
  {\bibinfo  {journal} {Eur. J. Org. Chem.}\ }\textbf {\bibinfo {volume}
  {2009}},\ \bibinfo {pages} {593} (\bibinfo {year} {2009})}\BibitemShut
  {NoStop}%
\bibitem [{\citenamefont {Lincoln}\ and\ \citenamefont
  {Joyce}(2009)}]{Lincoln2009}%
  \BibitemOpen
  \bibfield  {author} {\bibinfo {author} {\bibfnamefont {T.~A.}\ \bibnamefont
  {Lincoln}}\ and\ \bibinfo {author} {\bibfnamefont {G.~F.}\ \bibnamefont
  {Joyce}},\ }\href {\doibase 10.1126/science.1167856} {\bibfield  {journal}
  {\bibinfo  {journal} {Science}\ }\textbf {\bibinfo {volume} {323}},\ \bibinfo
  {pages} {1229} (\bibinfo {year} {2009})}\BibitemShut {NoStop}%
\bibitem [{\citenamefont {Esposito}\ and\ \citenamefont {Van~den
  Broeck}(2011)}]{Esposito2011}%
  \BibitemOpen
  \bibfield  {author} {\bibinfo {author} {\bibfnamefont {M.}~\bibnamefont
  {Esposito}}\ and\ \bibinfo {author} {\bibfnamefont {C.}~\bibnamefont {Van~den
  Broeck}},\ }\href@noop {} {\bibfield  {journal} {\bibinfo  {journal}
  {Europhys. Lett.}\ }\textbf {\bibinfo {volume} {95}},\ \bibinfo {pages}
  {40004} (\bibinfo {year} {2011})}\BibitemShut {NoStop}%
\bibitem [{\citenamefont {Parrondo}\ \emph {et~al.}(2015)\citenamefont
  {Parrondo}, \citenamefont {Horrowitz},\ and\ \citenamefont
  {Sagawa}}]{Parrondo2015}%
  \BibitemOpen
  \bibfield  {author} {\bibinfo {author} {\bibfnamefont {J.~M.}\ \bibnamefont
  {Parrondo}}, \bibinfo {author} {\bibfnamefont {J.~M.}\ \bibnamefont
  {Horrowitz}}, \ and\ \bibinfo {author} {\bibfnamefont {T.}~\bibnamefont
  {Sagawa}},\ }\href@noop {} {\bibfield  {journal} {\bibinfo  {journal} {Nat.
  Phys.}\ }\textbf {\bibinfo {volume} {11}},\ \bibinfo {pages} {131} (\bibinfo
  {year} {2015})}\BibitemShut {NoStop}%
\bibitem [{\citenamefont {Horowitz}\ \emph {et~al.}(2013)\citenamefont
  {Horowitz}, \citenamefont {Sagawa},\ and\ \citenamefont
  {Parrondo}}]{Horowitz2013}%
  \BibitemOpen
  \bibfield  {author} {\bibinfo {author} {\bibfnamefont {J.~M.}\ \bibnamefont
  {Horowitz}}, \bibinfo {author} {\bibfnamefont {T.}~\bibnamefont {Sagawa}}, \
  and\ \bibinfo {author} {\bibfnamefont {J.~M.~R.}\ \bibnamefont {Parrondo}},\
  }\href {\doibase 10.1103/PhysRevLett.111.010602} {\bibfield  {journal}
  {\bibinfo  {journal} {Phys. Rev. Lett.}\ }\textbf {\bibinfo {volume} {111}},\
  \bibinfo {pages} {010602} (\bibinfo {year} {2013})}\BibitemShut {NoStop}%
\bibitem [{\citenamefont {Ouldridge}\ \emph {et~al.}(ress)\citenamefont
  {Ouldridge}, \citenamefont {Govern},\ and\ \citenamefont
  {Wolde}}]{Ouldridge:2015vi}%
  \BibitemOpen
  \bibfield  {author} {\bibinfo {author} {\bibfnamefont {T.~E.}\ \bibnamefont
  {Ouldridge}}, \bibinfo {author} {\bibfnamefont {C.~C.}\ \bibnamefont
  {Govern}}, \ and\ \bibinfo {author} {\bibfnamefont {P.~R.}\ \bibnamefont
  {Wolde}},\ }\href@noop {} {\bibfield  {journal} {\bibinfo  {journal} {Phys.
  Rev. X}\ } (\bibinfo {year} {In Press})}\BibitemShut {NoStop}%
\bibitem [{\citenamefont {McGrath}\ \emph {et~al.}(2017)\citenamefont
  {McGrath}, \citenamefont {Jones}, \citenamefont {{ten Wolde}},\ and\
  \citenamefont {Ouldridge}}]{McGrath2016}%
  \BibitemOpen
  \bibfield  {author} {\bibinfo {author} {\bibfnamefont {T.}~\bibnamefont
  {McGrath}}, \bibinfo {author} {\bibfnamefont {N.~S.}\ \bibnamefont {Jones}},
  \bibinfo {author} {\bibfnamefont {P.~R.}\ \bibnamefont {{ten Wolde}}}, \ and\
  \bibinfo {author} {\bibfnamefont {T.~E.}\ \bibnamefont {Ouldridge}},\
  }\href@noop {} {\bibfield  {journal} {\bibinfo  {journal} {Phys. Rev. Lett.}\
  }\textbf {\bibinfo {volume} {118}},\ \bibinfo {pages} {028101} (\bibinfo
  {year} {2017})}\BibitemShut {NoStop}%
\bibitem [{\citenamefont {England}(2013)}]{England:2013ed}%
  \BibitemOpen
  \bibfield  {author} {\bibinfo {author} {\bibfnamefont {J.~L.}\ \bibnamefont
  {England}},\ }\href@noop {} {\bibfield  {journal} {\bibinfo  {journal} {J.
  Chem. Phys.}\ }\textbf {\bibinfo {volume} {139}},\ \bibinfo {pages} {121923}
  (\bibinfo {year} {2013})}\BibitemShut {NoStop}%
\bibitem [{SI()}]{SI}%
  \BibitemOpen
  \href@noop {} {\bibinfo  {journal} {{See Supplemental Material for additional derivations}}\ }\BibitemShut
  {NoStop}%
\bibitem [{\citenamefont {Huang}(1987)}]{Huang1987}%
  \BibitemOpen
\bibfield  {journal} {  }\bibfield  {author} {\bibinfo {author} {\bibfnamefont
  {K.}~\bibnamefont {Huang}},\ }\href@noop {} {\emph {\bibinfo {title}
  {Statistical Mechanics, Second Edition}}}\ (\bibinfo  {publisher} {John Wiley
  {\&} Sons, Inc.},\ \bibinfo {address} {New York},\ \bibinfo {year}
  {1987})\BibitemShut {NoStop}%
\bibitem [{\citenamefont {Bennett}(1982)}]{Bennett1982}%
  \BibitemOpen
  \bibfield  {author} {\bibinfo {author} {\bibfnamefont {C.~H.}\ \bibnamefont
  {Bennett}},\ }\href@noop {} {\bibfield  {journal} {\bibinfo  {journal} {Int.
  J. Theor. Phys.}\ }\textbf {\bibinfo {volume} {21}},\ \bibinfo {pages} {905}
  (\bibinfo {year} {1982})}\BibitemShut {NoStop}%
\bibitem [{\citenamefont {Andrieux}\ and\ \citenamefont
  {Gaspard}(2013)}]{Andrieux2013}%
  \BibitemOpen
  \bibfield  {author} {\bibinfo {author} {\bibfnamefont {D.}~\bibnamefont
  {Andrieux}}\ and\ \bibinfo {author} {\bibfnamefont {P.}~\bibnamefont
  {Gaspard}},\ }\href@noop {} {\bibfield  {journal} {\bibinfo  {journal}
  {Europhys. Lett.}\ }\textbf {\bibinfo {volume} {103}},\ \bibinfo {pages}
  {30004} (\bibinfo {year} {2013})}\BibitemShut {NoStop}%
\bibitem [{\citenamefont {Gaspard}\ and\ \citenamefont
  {Andrieux}(2014)}]{Gaspard2014}%
  \BibitemOpen
  \bibfield  {author} {\bibinfo {author} {\bibfnamefont {P.}~\bibnamefont
  {Gaspard}}\ and\ \bibinfo {author} {\bibfnamefont {D.}~\bibnamefont
  {Andrieux}},\ }\href@noop {} {\bibfield  {journal} {\bibinfo  {journal} {J.
  Chem. Phys.}\ }\textbf {\bibinfo {volume} {141}},\ \bibinfo {pages} {044908}
  (\bibinfo {year} {2014})}\BibitemShut {NoStop}%
\bibitem [{\citenamefont {Gaspard}(2015)}]{Gaspard2015}%
  \BibitemOpen
  \bibfield  {author} {\bibinfo {author} {\bibfnamefont {P.}~\bibnamefont
  {Gaspard}},\ }\href@noop {} {\bibfield  {journal} {\bibinfo  {journal} {Eur.
  Phys. J. Special Topics}\ }\textbf {\bibinfo {volume} {224}},\ \bibinfo
  {pages} {825} (\bibinfo {year} {2015})}\BibitemShut {NoStop}%
\bibitem [{\citenamefont {Gaspard}(2016)}]{Gaspard2016}%
  \BibitemOpen
  \bibfield  {author} {\bibinfo {author} {\bibfnamefont {P.}~\bibnamefont
  {Gaspard}},\ }\href {\doibase 10.1007/s10955-016-1532-x} {\bibfield
  {journal} {\bibinfo  {journal} {J. Stat. Phys.}\ }\textbf {\bibinfo {volume}
  {164}},\ \bibinfo {pages} {17} (\bibinfo {year} {2016})}\BibitemShut
  {NoStop}%
\bibitem [{\citenamefont {Wilber}\ \emph {et~al.}(2007)\citenamefont {Wilber},
  \citenamefont {Doye}, \citenamefont {Louis}, \citenamefont {Noya},
  \citenamefont {Miller},\ and\ \citenamefont {Wong}}]{Wilber2007}%
  \BibitemOpen
  \bibfield  {author} {\bibinfo {author} {\bibfnamefont {A.~W.}\ \bibnamefont
  {Wilber}}, \bibinfo {author} {\bibfnamefont {J.~P.~K.}\ \bibnamefont {Doye}},
  \bibinfo {author} {\bibfnamefont {A.~A.}\ \bibnamefont {Louis}}, \bibinfo
  {author} {\bibfnamefont {E.~G.}\ \bibnamefont {Noya}}, \bibinfo {author}
  {\bibfnamefont {M.~A.}\ \bibnamefont {Miller}}, \ and\ \bibinfo {author}
  {\bibfnamefont {P.}~\bibnamefont {Wong}},\ }\href {\doibase
  10.1063/1.2759922} {\bibfield  {journal} {\bibinfo  {journal} {J. Chem.
  Phys.}\ }\textbf {\bibinfo {volume} {127}},\ \bibinfo {eid} {085106}
  (\bibinfo {year} {2007})}\BibitemShut {NoStop}%
\bibitem [{\citenamefont {Reinhardt}\ and\ \citenamefont
  {Frenkel}(2014)}]{Reinhardt2014}%
  \BibitemOpen
  \bibfield  {author} {\bibinfo {author} {\bibfnamefont {A.}~\bibnamefont
  {Reinhardt}}\ and\ \bibinfo {author} {\bibfnamefont {D.}~\bibnamefont
  {Frenkel}},\ }\href@noop {} {\bibfield  {journal} {\bibinfo  {journal} {Phys.
  Rev. Lett.}\ }\textbf {\bibinfo {volume} {112}},\ \bibinfo {pages} {238103}
  (\bibinfo {year} {2014})}\BibitemShut {NoStop}%
\bibitem [{\citenamefont {Hopfield}(1974)}]{Hopfield:1974ij}%
  \BibitemOpen
  \bibfield  {author} {\bibinfo {author} {\bibfnamefont {J.~J.}\ \bibnamefont
  {Hopfield}},\ }\href@noop {} {\bibfield  {journal} {\bibinfo  {journal}
  {Proc. Nat. Acad. Sci. USA}\ }\textbf {\bibinfo {volume} {71}},\ \bibinfo
  {pages} {4135} (\bibinfo {year} {1974})}\BibitemShut {NoStop}%
\end{thebibliography}
\end{document}